\RequirePackage{fix-cm}

\documentclass[twocolumn,epjc3]{svjour3}  

\usepackage{cite}
\usepackage{float}
\usepackage{booktabs}
\usepackage{xspace}
\usepackage{flushend}
\usepackage{slashed}
\usepackage{cuted}
\usepackage{multirow}
\usepackage{dcolumn}
\usepackage{amssymb}
\usepackage{cancel}
\usepackage{verbatim}
\usepackage[utf8]{inputenc}
\usepackage{xcolor}
\usepackage{tabularx}
\usepackage{amsmath}
\usepackage{mathtools}
\usepackage{mathrsfs}
\usepackage{array}
\newcolumntype{L}[1]{>{\raggedright\let\newline\\\arraybackslash\hspace{0pt}}m{#1}}
\newcolumntype{C}[1]{>{\centering\let\newline\\\arraybackslash\hspace{0pt}}m{#1}}
\newcolumntype{R}[1]{>{\raggedleft\let\newline\\\arraybackslash\hspace{0pt}}m{#1}}

\usepackage{eso-pic}
\newcommand{\reportnum}[2]{
  \AddToShipoutPictureBG*{%
    \AtPageUpperLeft{%
      \hspace{0.79\paperwidth}%
      \raisebox{#1\baselineskip}{%
        \makebox[0pt][l]{\textnormal{#2}}
  }}}%
}

\allowdisplaybreaks

\newcommand{\fett}[1]{\boldsymbol{#1}}

\newcommand{\be}{\begin{equation}}
\newcommand{\ee}{\end{equation}}

\definecolor{darkred}{rgb}{0.5,0,0}
\definecolor{darkgreen}{rgb}{0,0.5,0}
\definecolor{darkblue}{rgb}{0,0,0.5}

\usepackage{hyperref}
\hypersetup{ colorlinks,
linkcolor=darkblue,
filecolor=darkgreen,
urlcolor=darkred,
citecolor=darkblue }

\newcommand{\inspire}[1]{\href{http://inspirehep.net/search?p=find+J+#1}
 {{\color{black}[{\color{blue} {\small in}SPIRE}]}}}
\newcommand{\book}[1]{\href{http://inspirehep.net/search?p=#1}
 {{\color{black}[{\color{blue} {\small in}SPIRE}]}}}

\newcommand{\inspired}[1]{\href{http://inspirehep.net/search?p=#1}
 {{\color{black}[{\color{blue} {\small in}SPIRE}]}}}

\newcommand{\CLASS}{\textsc{class}}

\journalname{Eur. Phys. J. C}

\begin{document}

\reportnum{-3}{CPPC-2021-12}

\title{Weaker yet again: mass spectrum-consistent cosmological constraints on the neutrino lifetime}

\author{Joe Zhiyu Chen\thanksref{addr1,e1}
	\and
	Isabel M.~Oldengott\thanksref{addr2,e2}
	\and
	Giovanni Pierobon\thanksref{addr1,e3}
	\and
	Yvonne~Y.~Y.~Wong\thanksref{addr1,e4}
}

\thankstext{e1}{e-mail: zhiyu.chen@unsw.edu.au}
\thankstext{e2}{e-mail: isabel.oldengott@uclouvain.be}
\thankstext{e3}{e-mail: g.pierobon@unsw.edu.au}
\thankstext{e4}{e-mail: yvonne.y.wong@unsw.edu.au}

\institute{Sydney Consortium for Particle Physics  and Cosmology, School of Physics, The University of New South Wales, Sydney NSW 2052, Australia\label{addr1}
	\and
		Centre for Cosmology, Particle Physics and Phenomenology, Universit\'{e} catholique de Louvain, Chemin du cyclotron, 2, Louvain-la-Neuve B-1348, Belgium\label{addr2}
}

\date{\today}

\maketitle

\begin{abstract}
We consider invisible neutrino decay $\nu_H \to \nu_l + \phi$ in the ultra-relativistic limit and compute the  neutrino anisotropy loss rate relevant for the cosmic microwave background (CMB) anisotropies.  Improving on our previous work which assumed {\it massless} $\nu_l$ and $\phi$, we reinstate in this work the daughter neutrino mass $m_{\nu l}$ in a manner consistent with the experimentally determined neutrino mass splittings.  We find that a nonzero $m_{\nu l}$ introduces a {\it  new} phase space  factor in  the loss rate $\Gamma_{\rm T}$ proportional to $(\Delta m_\nu^2/m_{\nu_H}^2)^2$ in the limit 
of a small squared mass gap  between the parent and daughter neutrinos, i.e., $\Gamma_{\rm T} \sim (\Delta m_\nu^2/m_{\nu H}^2)^2 (m_{\nu H}/E_\nu )^5 (1/\tau_0)$, where $\tau_0$ is the $\nu_H$ rest-frame lifetime.  Using a  general form of this result, we update the  limit on $\tau_0$ using the  Planck 2018 CMB data.
We find that for a parent neutrino of mass $m_{\nu H} \lesssim 0.1$~eV, the new phase space factor {\it weakens} the constraint on its lifetime by up to a  factor of 50 if $\Delta m_\nu^2$ corresponds to the atmospheric mass gap and up to $10^{5}$ if the solar mass gap, in comparison with na\"{\i}ve estimates that assume $m_{\nu l}=0$.   The revised  constraints  are (i)~$\tau^0 \gtrsim (6 \to 10) \times 10^5$~s and $\tau^0 \gtrsim (400 \to 500)$~s if only one neutrino decays to a daughter neutrino separated by, respectively, the atmospheric and the solar mass gap,    and (ii)~$\tau^0 \gtrsim (2 \to 6) \times 10^7$~s  in the case of two decay channels with one near-common atmospheric mass gap.  In contrast to previous, na\"{\i}ve limits which scale as $m_{\nu H}^5$, these mass spectrum-consistent $\tau_0$ constraints  are remarkably independent of the parent mass and  open up a swath of parameter space within the projected reach of IceCube and other  neutrino telescopes in the next two decades.
\end{abstract}




\section{Introduction}

Precision measurements of the cosmic microwave background (CMB) anisotropies and the large-scale matter distribution have in the past two decades yielded some of the tightest constraints on particle physics.  The most widely known amongst these are upper limits on the neutrino mass sum, which at $\sum m_\nu \lesssim 0.12$~eV for a one-parameter extension of the standard $\Lambda$CDM model~\cite{Planck:2018vyg},%
\footnote{95\% credible limit derived from a 7-parameter fit to the data combination Planck TT,TE,EE+lowE+lensing+BAO~\cite{Planck:2018vyg}.}
supersedes at face value current laboratory $\beta$-decay end-point spectrum measurements by a factor of 30~\cite{Aker:2019uuj}.  Of lesser prominence albeit equal significance are CMB constraints on the neutrino lifetime.

Relativistic invisible neutrino decay occurring on time scales of CMB formation leave an interesting signature on the CMB  anisotropies.  In the cosmological context, Standard-Model neutrinos decouple at temperatures of  $T \sim 1$~MeV and free-stream thereafter.  Free-streaming could be inhibited, however, if non-standard collisional processes such as neutrino decay and especially its concurrent inverse decay---the latter of which is kinematically feasible only in the limit of an ultra-relativistic neutrino ensemble---should become efficient at $T \lesssim 1$~MeV.  At CMB formation times, the absence of neutrino free-streaming prevents the transfer of power from the monopole (density) and dipole (velocity divergence) fluctuations of the neutrino fluid to higher multipole moments (e.g., anisotropic stress)~\cite{Hannestad:2004qu}.
  Such a suppression of higher-order anisotropies manifests itself predominantly in the CMB primary  anisotropies as an enhancement of power at multipoles $\ell \gtrsim 200$  in the CMB temperature power spectrum, which can be exploited to place constraints on the non-standard interaction(s) responsible for the phenomenon~\cite{Hannestad:2005ex,Basboll:2008fx,Cyr-Racine:2013jua,Archidiacono:2013dua,Escudero:2019gfk,Barenboim:2020vrr,Lancaster:2017ksf,Oldengott:2014qra,Oldengott:2017fhy,Kreisch:2019yzn,Forastieri:2019cuf,Barenboim:2019tux,Venzor:2022hql}.

Following this line of argument, measurements of the Planck mission (2018 data release)
currently constrain the rest-frame neutrino lifetime to $\tau_0 \gtrsim (4 \times 10^5 \to 4 \times 10^6)\ (m_\nu/50\,  {\rm meV})^5$~s, assuming invisible decay of a parent neutrino  not exceeding $m_{\nu} \simeq 0.2$~eV in mass into a massless particles~\cite{Barenboim:2020vrr}.%
 \footnote{It is also possible to invoke invisible neutrino decay in the non-relativistic limit as a means to relax cosmological neutrino mass bounds.  See, e.g., Refs.~\cite{Lorenz:2021alz,Abellan:2021rfq} for recent studies.  The rest-frame lifetimes required are typically of order $0.01 \to 0.1$ of the universe's age, leading to a phenomenology that differs completely from the ultra-relativistic case considered in this work.}
This lower bound is to be compared with laboratory limits obtained from analyses of the neutrino disappearance rate in JUNO and KamLAND+JUNO,  $\tau_0/m_\nu > 7.5 \times 10^{-11}$~(s/eV)  and 
$\tau_0/m_\nu >1.1 \times 10^{-9}$~(s/eV)  for a normal and an inverted mass ordering respectively~\cite{Porto-Silva:2020gma}, as well as astrophysical limits from solar neutrinos $\tau_0/m_\nu \gtrsim 10^{-5} \to 10^{-3}$~(s/eV)~\cite{SNO:2018pvg,Funcke:2019grs}, SN1987A $\tau_0\gtrsim (1 \to 10)$~s~\cite{Kachelriess:2000qc}, and IceCube $\tau_0/m_\nu \gtrsim 2.4 \times 10^3$~(s/eV)~\cite{Song:2020nfh}.

Central to the extraction of $\tau_0$ from cosmological data is the neutrino anisotropy loss rate or transport rate $\Gamma_{\rm T}$, particularly how this rate relates to $\tau_0$ itself and other fundamental properties of the particles participating in the interaction.  Some of us have recently considered this issue via a rigorous computation of the transport rate from the full Boltzmann hierarchies incorporating the two-body decay  $\nu_H \leftrightarrow \nu_l + \phi$ and its inverse process induced by a Yukawa interaction~\cite{Barenboim:2020vrr}.  
For decay into {\it massless} daughter particles, Ref.~\cite{Barenboim:2020vrr} finds a transport rate $\Gamma_{\rm T} \sim (m_\nu/E_\nu )^5 (1/\tau_0)$ under the assumption of linear response.  This result is in contrast to the rate $\Gamma_{\rm T} \sim (m_\nu/E_\nu )^3 (1/\tau_0)$ employed in previous analyses~\cite{Hannestad:2005ex,Basboll:2008fx,Archidiacono:2013dua,Escudero:2019gfk}, which is based upon a heuristic but (as we shall show) incorrect random walk argument.

 While representing a considerable step forward, the result  of Ref.~\cite{Barenboim:2020vrr} is nonetheless incomplete in that the mass difference between the parent neutrino~$\nu_H$ and daughter neutrino~$\nu_l$  must respect the neutrino mass spectra allowed by neutrino oscillations experiments.  Global three-flavour analyses of oscillation data currently find the squared mass differences to be~\cite{deSalas:2020pgw,Esteban:2020cvm,Capozzi:2017ipn}  
 \begin{eqnarray}
 \Delta m^2_{21} &=& 7.50^{+0.22}_{-0.20} \times 10^{-5}~{\rm eV}^2, \\
 |\Delta m^2_{31}|_{\rm N}& =&  2.55^{+0.02}_{-0.03} \times 10^{-3}~{\rm eV}^2,   \\ 
 |\Delta m^2_{31}|_{\rm I} &=& 2.45^{+0.02}_{-0.03} \times 10^{-3}~{\rm eV}^2, 
 \end{eqnarray}
where $|\Delta m^2_{31}|_{\rm N}$ applies to a normal mass ordering (NO; $m_3 > m_2 > m_1$), and $|\Delta m^2_{31}|_{\rm I}$ to an inverted mass ordering (IO; $m_2 > m_1 > m_3$).
 Relative to the benchmark parent neutrino mass of $0.2$~eV, these squared mass differences are clearly small enough that the assumption of a massless daughter neutrino may become untenable and consequently alter the relationship between the transport rate $\Gamma_{\rm T}$ and the rest-frame decay rate $\tau_0$ away from $\Gamma_{\rm T} \sim (m_\nu/E_\nu )^5 (1/\tau_0)$.
 
In this work, we improve upon the original calculation of Ref.~\cite{Barenboim:2020vrr} to explicitly allow for a massive daughter neutrino $\nu_l$.  We furthermore extend the computation of the anisotropic stress (i.e., multipole $\ell=2$) damping rate to $\ell > 2$ multipole moments---for compatibility with the neutrino Boltzmann hierarchy~\cite{Ma:1995ey} solved in the CMB code \CLASS{}~\cite{Blas:2011rf}---and  put the generalisation of the calculation to other interaction structures (besides Yukawa) on firmer grounds.
  We find that, in the presence of a massive $\nu_l$, a consistent transport rate $\Gamma_{\rm T}$ is suppressed by an extra phase space factor proportional to $(\Delta m_\nu^2/m_\nu^2)^2$ in the limit of a small squared mass difference between the parent and daughter neutrinos, i.e.,  $\Gamma_{\rm T} \sim (\Delta m_\nu^2/m_\nu^2)^2 (m_\nu/E_\nu )^5 (1/\tau_0)$,
  and this factor is {\it in addition to} the phase space suppression ($\sim \Delta m_\nu^2/m_\nu^2$) already inherent in the lifetime $\tau_0$.
  
Using a general version of this result valid for all $\Delta m_\nu^2$, we fit the Planck 2018 CMB data to derive a new set of cosmological  constraints on the neutrino lifetime as functions of the lightest neutrino mass (i.e., $m_1$ in NO and $m_3$ in IO) that are consistent with mass difference measurements.    
 Readers interested in a quick number can jump directly to  Fig.~\ref{fig:scenarioA} where the new constraints on $\tau_0$ are presented.   Here, we note that for a parent neutrino of  $m_\nu \lesssim 0.1$~eV, the new phase space suppression factor weakens the CMB constraint on its lifetime by as much as a  factor of 50 if $\Delta m_\nu^2$ corresponds to the atmospheric mass gap and by five orders of magnitude if the solar mass gap, compared with na\"{\i}ve limits that do not account for the daughter neutrino mass.

The paper is organised as follows.  We begin with a description of the physical system in Sect.~\ref{sec:physicalsystem}. We then introduce the basic equations of motion for the individual $\nu_H$, $\nu_l$ and $\phi$ component in Sect.~\ref{sec:basic}, before condensing them into a single set of effective equations of motion for the combined neutrino+$\phi$ system in Sect.~\ref{sec:effectiveEoM}.  Using standard statistical inference techniques presented in Sect.~\ref{sec:mcmc}, we derive in Sect.~\ref{sec:results} new CMB constraints on the neutrino lifetime that are consistent with the experimentally determined neutrino mass splittings.   We conclude in Sect.~\ref{sec:conclusions}.     Four appendices contain respectively  full expressions of the collision integrals (\ref{sec:individualcollisionintegral}), computational details of the $\ell \geq 2$ effective damping rates (\ref{sec:collisionintegral}), a revised random walk picture  explaining the $\propto (m_\nu/E_\nu)^5$ dependence of the transport rate and why the old $\propto (m_\nu/E_\nu)^3$ rate is incorrect (\ref{sec:heuristic}), and summary statistics of our parameter estimation analyses (\ref{sec:parameter}).


\section{The physical system}
\label{sec:physicalsystem}

We consider the two-body decay of a parent neutrino $\nu_H$  into a massive daughter neutrino $\nu_l$ and a massless particle $\phi$, whose transition probability is described by a Lorentz-invariant squared matrix element~$|{\cal M}|^2$.  Previously in Ref.~\cite{Barenboim:2020vrr} we had assumed~$\phi$ to be a scalar and the decay described by a Yukawa interaction,  motivated by Majoron models~\cite{Gelmini:1980re,Chikashige:1980ui,Schechter:1981cv}  in which a new Goldstone boson of a spontaneously broken global $U(1)_{B-L}$ symmetry couples to neutrinos.  Another possibility arises within gauged $L_i -L_j$ models, where a new vector boson  plays the role of the massless  state~\cite{Dror:2020fbh,Ekhterachian:2021rkx,Chen:2021qaf,Alonso-Alvarez:2021pgy}. Models of neutrinos decaying on cosmological timescales can be found in Refs. \cite{Gelmini:1983ea,Joshipura:1992vn,Akhmedov:1995wd,Escudero:2020ped}. Here, in order to keep our analysis as general as possible, apart from the quantum statistical requirement that $\phi$ must be boson, we shall leave its nature and the nature of the interaction unspecified.

Because cosmological observables do not measure neutrino spins, it is not possible to define a reference direction in the decaying neutrino's rest-frame.  Thus, for an ensemble of neutrinos with randomly aligned spins, we can effectively consider the decay to be isotropic in the decaying neutrino's rest-frame, with a decay rate $\Gamma_{\rm dec}^0 \equiv 1/\tau_0$ related to the squared matrix element via
\begin{equation}
	\Gamma^0_{\rm dec} = \frac{ \Delta (m_{\nu H},m_{\nu l},m_\phi)}{16 \pi  m_{\nu H}^3} |{\cal M}|^2,
	\label{eq:restframerate1}
\end{equation}
where 
\begin{equation}
	\Delta(x,y,z) 
\equiv \sqrt{\left[x^2-\left(y+z\right)^2 \right] \left[x^2-\left(y-z\right)^2 \right]}\\
\end{equation}
is the K\"{a}ll\'{e}n function, and it is understood that $|{\cal M}|^2$ needs to be evaluated at particle energies and  momenta consistent with conservation laws.

In the cosmological context, the decay process can be expected to come into effect around scale factors $a \sim a_{\rm dec}$, where $a_{\rm dec}$ is defined via  $\Gamma_{\rm dec} (a_{\rm dec})= H(a_{\rm dec})$, with $\Gamma_{\rm dec}=\left\langle m_{\nu H}/E_{\nu H} \right\rangle  \Gamma_{\rm dec}^0$  the ensemble-averaged Lorentz-boosted decay rate.
We are interested in the decay of $\nu_H$ occurring (i)~up to the time of recombination but substantially after neutrino decoupling, i.e., $10^{-10} \ll a_{\rm dec} \lesssim 10^{-3}$, and (ii) while the bulk of the $\nu_H$ ensemble is ultra-relativistic.
 Two observations are in order:
\begin{enumerate}
\item The ultra-relativistic requirement guarantees that the inverse decay process $\nu_l + \phi \to \nu_H$ is kinematically allowed and kicks in on roughly the same time scale as $\nu_H$ decay, 
enabling $\nu_H$, $\nu_l$, and $ \phi$ to form a coupled system whose combined anisotropic stress can be suppressed through collisional damping.
The same requirement also effectively limits the parent neutrino mass to the ballpark  $m_{\nu H} \lesssim 3 T_{\nu} (a_{\rm rec})\sim  3 \times (4/11)^{1/3} T_{\gamma} (a_{\rm rec}) \ \sim 0.2$~eV, if the phenomenology of anisotropic stress loss is to remain unambiguously applicable up to the  recombination era when the temperature of the photon bath drops to $T_\gamma(a_{\rm rec}) \sim 0.1$~eV.

\item Since decay and inverse decay occur by construction  only after neutrino decoupling, our scenario does not generate an excess of radiation.  Rather, the $(\nu_H,\nu_l,\phi)$-system merely shares the energy that was in the original $\nu_H$ and $\nu_l$ populations.
Consequently, in the time frame of interest cosmology is unaffected at the background level, and the  phenomenology of the $(\nu_H,\nu_l,\phi)$-system 
{\it vis-\`{a}-vis} the CMB primary anisotropies is at leading order limited  to anisotropic stress loss.

\end{enumerate}
 These observations tell us that, as far as the CMB primary anisotropies are concerned, 
 the $(\nu_H,\nu_l,\phi)$-system can be modelled at leading order as a single massless fluid whose energy density is equivalent to that of the original $\nu_H$ and $\nu_l$ populations at $a <a_{\rm dec}$.  What remains to be specified is the rate at which the fluid's anisotropic stress and higher-order multipole moments are lost via decay and inverse decay; this is the subject of Sects.~\ref{sec:basic} and~\ref{sec:effectiveEoM}.


\section{Basic equations of motion}
\label{sec:basic}

A complete set of evolution equations for the individual $\nu_H$, $\nu_l$, and $\phi$ phase space density under the influence of gravity and decay/inverse decay has been presented in Ref.~\cite{Barenboim:2020vrr}.  We briefly describe the relevant ones below.

We work in the synchronous gauge defined by the invariant 
${\rm d}s^2 = a^2(\tau) \left[- {\rm d} \tau^2 + (\delta_{ij} + h_{ij}) {\rm d} x^i {\rm d} x^j \right]$,
and split the Fourier-transformed phase space density of species $i$ into a homogeneous and isotropic component and a perturbed part, i.e.,
$f_i(\mathbf{k},\mathbf{q},\tau) = \bar{f}_i(|\mathbf{q}|,\tau) + F_i (\mathbf{k},\mathbf{q},\tau)$,
where $\tau$ is conformal time, $\mathbf{q}\equiv |\mathbf{q}| \hat{q}$  the comoving 3-momentum, and $\mathbf{k} \equiv |\mathbf{k}| \hat{k}$ is the Fourier wave vector conjugate to the comoving spatial coordinates $\mathbf{x}$ of the line element.

Because for the scenario at hand the $(\nu_H,\nu_l,\phi)$-system behaves at leading order essentially as a single massless fluid whose energy density remains constant in the time frame of interest, the evolution of the individual homogeneous phase space density $\bar{f}_i(|\mathbf{q}|,\tau)$ is not a main concern.  In fact, in Sect.~\ref{sec:effectiveEoM} we shall approximate all $\bar{f}_i$'s with  equilibrium distributions in order to derive closed-form expressions for the effective damping rates.

Of greater interest to us is the perturbed component $F_i (\mathbf{k},\mathbf{q},\tau)$.
Following standard practice~\cite{Ma:1995ey}, we decompose $F_i$ in terms of a Legendre series,
\begin{equation}
\begin{aligned}
		F_{i}(|\mathbf{k}|,|\mathbf{q}|,x) & \equiv \sum_{\ell=0}^{\infty}(-{\rm i})^{\ell} (2\ell+\!1) F_{i,\ell}(|\mathbf{k}|,|\mathbf{q}|) P_{\ell}(x), \\
		F_{i,\ell}(|\mathbf{k}|,|\mathbf{q}|) & \equiv \frac{{\rm i}^{\ell}}{2} \int_{-1}^{1} \mathrm{d}x \, F_{i}(|\mathbf{k}|,|\mathbf{q}|,x) P_{\ell}(x),
		\label{Legendre_decompositio0n}
		\end{aligned}
\end{equation}
where $x\equiv \hat{k}\cdot \hat{q}$, and $P_\ell(x)$ is a Legendre polynomial of degree $\ell$.  Then, the equations of motion for the Legendre moments ${F}_{i,\ell}$ can be written as
\begin{equation}
	\begin{aligned}
		\dot{F}_{i,0} & = -\frac{|\mathbf{q}| |\mathbf{k}|}{\epsilon_i} F_{i,1}+ \frac{1}{6} \frac{\partial  \bar{f}_{i}}{\partial \ln |\mathbf{q}|}\dot{h} + \left( \frac{{\rm d} f_i}{{\rm d} \tau}\right)_{C,0}^{(1)} , \\
		\dot{F}_{i,1}&= \frac{|\mathbf{q}| |\mathbf{k}|}{\epsilon_i} \left(-\frac{2}{3} F_{i,2}+\frac{1}{3}  F_{i,0}\right) + \left( \frac{{\rm d} f_i}{{\rm d} \tau}\right)_{C,1}^{(1)},\\
		\dot{F}_{i,2} &= \frac{|\mathbf{q}| |\mathbf{k}|}{\epsilon_i} \left(-\frac{3}{5} F_{i,3}  +  \frac{2}{5} F_{i,1} \right)  \\
		& \qquad-  \frac{\partial  \bar{f}_{i}}{\partial \ln |\mathbf{q}|}  \left( \frac{2}{5} \dot{\eta}  +  \frac{1}{15} \dot{h} \right)  + \left( \frac{{\rm d} f_i}{{\rm d} \tau}\right)_{C,2}^{(1)} ,\\
		\dot{F}_{i,\ell>2} & =\, \frac{|\mathbf{k}|}{2\ell+1} \frac{|\mathbf{q}|}{\epsilon_i} \left[ \ell F_{i,\ell-1}  -  (\ell +  1)F_{i,\ell+1} \right]  \\
		& \qquad \qquad +  \left( \frac{{\rm d} f_i}{{\rm d} \tau}\right)_{C,\ell}^{(1)} .
	\end{aligned}
	\label{eq:hierarchyF}
\end{equation}
Here, $\cdot \equiv \partial/\partial \tau$, $\epsilon_i (|\mathbf{q}|)\equiv (|\mathbf{q}|^2 + a^2 m_i^2)^{1/2}$ is the comoving energy of the  particle species~$i$ of mass~$m_i$, $h \equiv \delta^{ij} h_{ij}(\mathbf{k},\tau)$ and $\eta = \eta(\mathbf{k},\tau) \equiv - k^i k^j h_{ij}/(4 k^2) + h/12$ are the trace and traceless components of $h_{ij}$ in Fourier space respectively, and 
$\left( {\rm d} f_i/{\rm d} \tau \right)_{C,\ell}^{(1)}$ represents the $\ell$th Legendre moment of the linear-order Boltzmann collision integral.

The linear-order decay/inverse decay collision integrals  have been derived in Ref.~\cite{Barenboim:2020vrr}
assuming a Yukawa interaction.  However, following the arguments of Sect.~\ref{sec:physicalsystem} and using the fact that the squared matrix element is Lorentz-invariant,  it is straightforward to generalise the integrals of Ref.~\cite{Barenboim:2020vrr} to any interaction structure.  The same argument also allows us to recast the collision integrals in favour of the rest-frame decay rate $\Gamma_{\rm dec}^0$ as a fundamental parameter (over the squared matrix element $|{\cal M}|^2$).  The interested reader can find the full expressions in Eqs.~\eqref{eq:C1_nuH}--\eqref{eq:C1_phi} in \ref{sec:individualcollisionintegral}.

The equations of motion~\eqref{eq:hierarchyF} together with the collision integrals~\eqref{eq:C1_nuH}--\eqref{eq:C1_phi} can be programmed directly into and solved by a CMB code such as~\CLASS{}~\cite{Blas:2011rf}.
In practice, however, numerical solutions particularly  in the ultra-relativistic limit can be highly non-trivial, because of the vastly different time scales at play: the basic decay time scale $1/\Gamma_{\rm dec}$ and the emergent transport time scale $1/\Gamma_{\rm T} \sim (E_\nu/m_\nu)^4 (1/\Gamma_{\rm dec})$.  In Sect.~\ref{sec:effectiveEoM} we show how these equations of motion can be condensed in the ultra-relativistic limit  into a single set of equations for the combined $(\nu_H,\nu_l,\phi)$-system under certain reasonable simplifying assumptions.


\section{Effective equations of motion for the neutrino+\texorpdfstring{$\phi$}{phi} system}
\label{sec:effectiveEoM}

We are interested in the evolution of perturbations in the combined $(\nu_H,\nu_l,\phi)$-system in the ultra-relativistic limit.  To this end we define the integrated Legendre moment,
\begin{equation}
\begin{aligned}
		{\cal F}_\ell(|\mathbf{k}|)  \equiv \, &  \frac{(2 \pi^2 a^4)^{-1}}{\bar{\rho}_{\nu\phi}}  \sum_{i=\nu_H,\nu_l,\phi}  g_i \\
		&\quad \quad \times \int {\rm d} |\mathbf{q}|\; |\mathbf{q}|^2 \epsilon_i \left(\frac{|\mathbf{q}|}{\epsilon_i} \right)^\ell  F_{i,\ell} (|\mathbf{q}|),
		\label{eq:integratedF}
		\end{aligned}
			\end{equation}
where $g_i$ is the internal degree of freedom of the $i$th particle species, and  $\bar{\rho}_{\nu\phi}$ represents the total background energy density of the $(\nu_H,\nu_l,\phi)$-system which we take to be ultra-relativistic, i.e., $\bar{\rho}_{\nu \phi} \propto a^{-4}$, and equal to the energy density in the pre-decay $\nu_H$ and $\nu_l$ populations.

Summing and integrating the equations of motion, Eq.~\eqref{eq:hierarchyF}, as per the definition~\eqref{eq:integratedF} and taking the ultra-relativistic limit everywhere {\it except}  the collision terms, we obtain
\begin{equation}
\begin{aligned}
	\dot{\cal F}_0 & = - |\mathbf{k}| {\cal F}_1 - \frac{2}{3} \dot{h}, \\
	\dot{\cal F}_1  & = \frac{|\mathbf{k}|}{3} \left( {\cal F}_0 -  2{\cal F}_2 \right),\\
	\dot{\cal F}_{2} &= \frac{|\mathbf{k}|}{5}  \left( 2 {\cal F}_1 - 3  {\cal F}_3 \right)+ \frac{4}{15} \dot{h} + \frac{8}{5} \dot{\eta} +  \left( \frac{{\rm d} {\cal F}}{{\rm d} \tau}\right)_{C,2} , \\
	\dot{\cal F}_{\ell>2} & =\, \frac{|\mathbf{k}|}{2\ell+1}\left[ \ell {\cal F}_{\ell-1} - (\ell+1){\cal F}_{\ell+1} \right] +  \left( \frac{{\rm d} {\cal F}}{{\rm d} \tau}\right)_{C,\ell} ,
	\label{eq:hierarchytotalF}
\end{aligned}
\end{equation}
where
\begin{equation}
	\begin{aligned}
	\left(\frac{\mathrm{d} {\cal F}}{\mathrm{d}\tau}\right)_{C,\ell} & \equiv\,  
		\frac{(2 \pi^2 a^4)^{-1}}{\bar{\rho}_{\nu\phi}}
		\sum_{i=\nu_H,\nu_l,\phi} g_i  \\
		&\times \int   {\rm d} |\mathbf{q}|\; |\mathbf{q}|^2 \epsilon_i  \left(\frac{|\mathbf{q}|}{\epsilon_i} \right)^\ell  \left(\frac{\mathrm{d} f_i}{\mathrm{d}\tau}\right)_{C,  \ell}^{(1)}  (|\mathbf{q}|)
		\label{eq:dpidt}
	\end{aligned}
\end{equation}
are the effective collision integrals constructed from the individual collision integrals~\eqref{eq:C1_nuH}--\eqref{eq:C1_phi}.
Note that the effective collision integrals for $\ell=0,1$ integrated moments must vanish exactly because of energy and momentum conservation in the $(\nu_H,\nu_l,\phi)$-system.


\subsection{Effective collision integrals}
\label{sec:coll}

Equations~\eqref{eq:hierarchytotalF} and~\eqref{eq:dpidt} are not yet in a closed form.  To put them into a closed form, several simplifying approximations to the effective collision integrals are in order~\cite{Barenboim:2020vrr}:
\begin{enumerate}
	\item We assume equilibrium Maxwell-Boltzmann statistics, meaning that the background phase space density of each particle species can be described at any instant by $\bar{f}_i(|\mathbf{q}|) = e^{-(\epsilon_i-\mu_i)/T_0}$, with a common comoving temperature $T_0$ and chemical potentials satisfying $\mu_{\nu H} = \mu_{\nu l} + \mu_\phi$.  Physically, this means we take the background equilibration rate $\Gamma_{\rm dec}$ to be much larger than the loss rates of the  anisotropic stress and higher-order anisotropies  we are interested to compute.  As demonstrated in Ref.~\cite{Barenboim:2020vrr}, relativistic $\nu_H$ decay and inverse decay attain a steady-state/quasi-equilibrium regime in which the above equilibrium conditions are reasonably well satisfied.

	\item We take the individual species' Legendre moments $F_{i,\ell}(|\mathbf{k}|,|\mathbf{q}|)$ to be separable functions of $|\mathbf{k}|$ and $|\mathbf{q}|$, given by~\cite{Oldengott:2017fhy}
	\begin{equation}
	F_{i,\ell} (|\mathbf{k}|,|\mathbf{q}|)  \simeq - \frac{1}{4} \frac{{\rm d} \bar{f}_i}{{\rm d} \ln |\mathbf{q}|}\, {\cal F}_{i,\ell} (|\mathbf{k}|),
	\label{eq:separable}
	\end{equation}
	where ${\cal F}_{i,\ell}$ is the analogue of the integrated Legendre moment ${\cal F}_{\ell}$ of Eq.~\eqref{eq:integratedF} but for the $i$th species alone. In the absence of collisions, this so-called separable {\it ansatz}~\eqref{eq:separable} is exact in the ultra-relativistic limit and reflects the fact that, for massless particles, gravity is achromatic and all $|\mathbf{q}|$-modes evolve in the same way. For collisional systems near equilibrium, i.e., where the momentum-dependence of the background phase space densities is largely constant in time, the {\it ansatz} also appears to capture the evolution of $ {\cal F}_{i,\ell\leq 2} (|\mathbf{k}|)$ well in the ultra-relativistic limit, across a range of $|\mathbf{k}|$-modes and redshifts~\cite{Oldengott:2017fhy}.
	
	\item In standard cosmology, all $\ell\geq2$ anisotropies are induced by gravity after a Fourier $\mathbf{k}$-mode enters the horizon.  Gravity induces the same perturbation contrast ${\cal F}_{i,\ell}$ in all particle species at the same point in space, which immediately leads us to our third approximation
	\begin{equation}
	{\cal F}_{\nu H,\ell} (|\mathbf{k}|) \simeq  {\cal F}_{ \nu l,\ell} (|\mathbf{k}|)  \simeq  {\cal F}_{\phi,\ell} (|\mathbf{k}|) \simeq {\cal F}_{\ell} (|\mathbf{k}|),
	\label{eq:sameperturbation}
	\end{equation}
where ${\cal F}_{\ell} (|\mathbf{k}|)$ is identically the $\ell$th integrated Legendre moment of Eq.~\eqref{eq:integratedF}.
\end{enumerate}
 Note that by making these reasonable assumptions, we have essentially recast the problem of establishing the effective transport rate into a relaxation time calculation in terms of the system's linear response to a small external perturbation.

Under the above approximations and assuming further that $m_\phi=0$, explicit evaluation of the effective collision integral~\eqref{eq:dpidt} yields 
\begin{equation}
\begin{aligned}
\label{eq:effectiveEoM}
\left(	\frac{{\rm d} {\cal F}}{{\rm d} \tau} \right)_{C,\ell} = \, & - \alpha_\ell \, a \, \tilde{\Gamma}_{\rm dec} \left(\frac{ a m_{\nu H}}{T_0} \right)^4 \\
& \quad \quad  \times \mathscr{F} \left(\frac{a m_{\nu H}}{T_0}\right)\,   \Phi\left(\frac{m_{\nu l}}{m_{\nu H}} \right) \,  {\cal F}_\ell.
\end{aligned}
\end{equation}
See \ref{sec:collisionintegral} for details of the derivation.
Here, the rate $\tilde{\Gamma}_{\rm{dec}}$ is approximately the boosted decay rate and is defined as in 
Ref.~\cite{Barenboim:2020vrr} via
\begin{equation}
\begin{aligned}
\label{eq:tildegamma}
\tilde{\Gamma}_{\rm dec} \equiv \, &  \frac{(4 \pi^2 a^4)^{-1}}{\bar{\rho}_{\nu\phi}} \, a m_{\nu H} T_0^3 \, g_{\nu H} \, e^{\, \mu_{\nu H}/T_0} \, \Gamma^0_{\rm dec} \\
= \, &  \frac{1}{12} \, \frac{(a^4 \bar{\rho}_{\nu H})}{(a^4 \bar{\rho}_{\nu \phi})} \, \frac{a m_{\nu H}}{T_0} \,  \Gamma^0_{\rm dec} \, ,
\end{aligned}
\end{equation}
where $\bar{\rho}_{\nu H} \equiv (3 g_{\nu H} T_0^4/a^4 \pi^2)  \, e^{\, \mu_{\nu H} / T_0}$ is the homogeneous, background energy density of $\nu_H$ in the ultra-relativistic limit.    

The dimensionless function $\mathscr{F}(x)$ in Eq.~\eqref{eq:effectiveEoM} is a decreasing function of $x$ given by
\begin{equation}
\label{eq:curlyF}
	\mathscr{F}(x) 
	 =  \frac{1}{2} e^{-x} \Big[ - 1 + x - e^x (x^2-2) \Gamma(0,x) \Big],
\end{equation}
where $\Gamma(0,x)$ denotes the incomplete gamma function.  For $ x \sim 10^{-10}  \to 0.1$, $\mathscr{F}(x)$ evaluates to  $\sim 20 \to 1$.  
Beyond $x \sim 1$, $\mathscr{F}(x)$ quickly drops to zero, reflecting the fact that 
after $\nu_H$ becomes non-relativistic, inverse decay must also shut down so that the combined neutrino+$\phi$ fluid reverts to a free-streaming system.  In order to bypass the incomplete gamma function in our numerical solution, we find it convenient to use the small-$x$ expansion
\begin{equation}
\begin{aligned}
	\mathscr{F}_{\rm sm}(x) & \simeq   -\frac{1}{2} + 2 x - x^2 -\frac{x^3}{9} + \frac{x^4}{96} - \frac{x^5}{900} \\ 
	& + \frac{x^6}{8640}  +\left(\gamma + \log x\right) \left( \frac{x^2}{2} -1 \right) + {\cal O}(x^7), 
				\end{aligned}
	\end{equation}
where $\gamma \simeq 0.57721$ is Euler-Mascheroni constant.  The expansion is accurate to $<7$\% for $x \leq 2$. For $x > 2$, we switch to the function $\mathscr{F}_{\rm lg}(x)  \simeq 2.4 \, e^{-x}/x^{1.5}$, which approximates Eq.~\eqref{eq:curlyF} to $<10$\% at $2 < x < 17$.

Relative to the result of Ref.~\cite{Barenboim:2020vrr}, the collision integral~\eqref{eq:effectiveEoM}  also contains two new elements.  Firstly, the phase space factor,
\begin{equation}
\label{eq:phasespace}
\Phi (y)  = \frac{1}{1 - y^2}  \left( 1 - y^4 + 4\, y^2 \ln y \right), 
\end{equation}
allows for the possibility of a nonzero $m_{\nu l}$ and takes on values $1 \to 0$ in the region $0 \leq y < 1$ (or, equivalently, $0 \leq m_{\nu l} < m_{\nu H}$).  In the limit of a small squared mass gap, i.e.,  $\Delta m_\nu^2 \equiv m_{\nu H}^2-m_{\nu l}^2 \ll m_{\nu H}^2$, it reduces to $\Phi(m_{\nu l}/m_{\nu H}) \to (1/3) (\Delta m_\nu^2/m_{\nu H}^2)^2$ and traces its origin to the fact that the momentum carried by each decay product in the rest frame of $\nu_H$ 
is in fact $\Delta m_\nu^2/(2 m_{\nu H})$, not $m_{\nu H}/2$ as in the case of massless decay products.  
Secondly, the  $\ell$-dependent coefficients $\alpha_\ell$ are given by the expression
\begin{equation}
\alpha_\ell = \frac{1}{32} \left(3 \ell^4 + 2 \ell^3 - 11 \ell^2 + 6 \ell \right)
\label{eq:alpha}
\end{equation}
for all $\ell \geq 0$. These coefficients increase with $\ell$, reflecting the fact that, 
 for a fixed decay opening angle, the decay and inverse decay processes become  more efficient at wiping out anisotropies  on progressively smaller angular scales.
 
Lastly, we highlight again that the effective collision rate in Eq.~\eqref{eq:effectiveEoM} has in total five powers of $a m_{\nu H}/T_0$, one of which is the Lorentz boost appearing in the boosted decay rate~\eqref{eq:tildegamma}.  This result is in drastic contrast with the much larger $\propto (a m_{\nu H}/T_0)^3$ rate used in previous works~\cite{Hannestad:2005ex,Basboll:2008fx,Archidiacono:2013dua,Escudero:2019gfk}, which we stress is {\it not} an {\it ab initio} result, but rather one based upon a heuristic random walk argument.  
\ref{sec:heuristic} shows that this heuristic argument is in fact incomplete, and that a consistent random walk picture accounting for all same-order effects must yield the much smaller $\propto (a m_{\nu H}/T_0)^5$ rate derived from first principles in this work.


\subsection{Three-state to two-state approximation}
\label{sec:threetwo}

The decay scenario presented thus far concerns only two of the three Standard-Model neutrino mass states. Without model-specific motivations to exclude one particular mass state from the decay interaction, one might generally expect all three  mass states to couple similarly at the Lagrangian level and participate in the decay/inverse decay process as parent particles and/or decay products.
Then, kinematics largely determine the decay rates, and we can deduce from the measured neutrino squared mass differences that the rest-frame decay rates $\Gamma^0_{i \to j}$ for  the processes $\nu_i \to \nu_j + \phi$ follow the patterns
\begin{equation}
\label{eq:hierarchy1}
\Gamma_{3 \to 1 }^0 \simeq \Gamma^0_{3 \to 2} \gg \Gamma^0_{2 \to 1}
\end{equation}
assuming a normal mass ordering,
and 
\begin{equation}
\label{eq:hierarchy2}
\Gamma_{2 \to 3 }^0 \simeq \Gamma^0_{1 \to 3} \gg \Gamma^0_{2 \to 1}
\end{equation}
assuming an inverted mass ordering.   
All IO  mass spectra and those NO mass spectra consistent with  $m_1 \gtrsim 0.03$~eV   satisfy the condition~\eqref{eq:hierarchy1} or~\eqref{eq:hierarchy2} at better than 10\%.  For $m_1 \lesssim 0.03$~eV in NO, the ratio $\Gamma^0_{2 \to 1}/\Gamma^0_{3\to 2}$
rises to $\sim 0.25$ while $\Gamma^0_{3 \to 1}/\Gamma^0_{3\to 2}$ approaches
$0.85$~\cite{Barenboim:2020vrr}. Thus, Eq.~\eqref{eq:hierarchy1} is still a reasonably well satisfied.

If the two larger (boosted) decay rates satisfy the equilibrium condition $\Gamma_{i \to j} \gtrsim H$, then, irrespective of the mass ordering, we can expect the phenomenology of the three-flavour system to be similar to the two-flavour case discussed in Sect.~\ref{sec:physicalsystem}, save for the understanding that it is now the $\nu_1$, $\nu_2$, $\nu_3$, and $\phi$ that form a single massless fluid system.  It remains to determine the effective collision integral and hence the $\ell \geq 2$ anisotropy damping rates, which we approximate as follows.

\paragraph{Normal mass ordering}  Here, we assume the two lightest mass states with masses $m_1$ and $m_2$ to be effectively degenerate and set $\Gamma^0_{2 \to 1} = 0$.  Then, following the recipe laid down in Ref.~\cite{Barenboim:2020vrr}, the equations of motion can be condensed by multiplying by a factor of 2 (i)~the $\nu_H$ and $\phi$ collision integrals~\eqref{eq:C1_nuH} and~\eqref{eq:C1_phi}, and (ii)~all momentum-integrated quantities pertaining to~$\nu_l$.  This immediately leads to an effective collision integral
\begin{equation}
\begin{aligned}
\left(\frac{\mathrm{d} {\cal F}}{\mathrm{d}\tau}\right)_{C,\ell} & = \, 2 \, \frac{(2 \pi^2 a^4)^{-1}}{\bar{\rho}_{\nu\phi}} 
\sum_{i=\nu_H,\nu_l,\phi} g_i \\
& \times \int {\rm d} |\mathbf{q}|\; |\mathbf{q}|^2 \epsilon_i \left(\frac{|\mathbf{q}|}{\epsilon_i} \right)^\ell \, \left(\frac{\mathrm{d} f_i}{\mathrm{d}\tau}\right)_{C,  \ell}^{(1)}  (|\mathbf{q}|) , \\
\label{eq:NHdpidt}
\end{aligned}
\end{equation}
 that differs from the original definition~\eqref{eq:dpidt} and hence Eq.~\eqref{eq:effectiveEoM} by an overall factor of~2.

\paragraph{Inverted mass ordering}  Here, we treat the two heaviest mass states as effectively degenerate and again set $\Gamma^0_{2 \to 1}=0$.  Then, multiplying by a factor of 2 (i)~the $\nu_l$ and $\phi$ collision integrals~\eqref{eq:C1_nul} and~\eqref{eq:C1_phi}, and (ii)~all momentum-integrated quantities of $\nu_H$, it is straightforward to show the same overall factor of 2 appears again in front of the effective collision integral as in the normal mass ordering case in Eq.~\eqref{eq:NHdpidt}.


\subsection{Effective temperature and density ratio}
\label{sec:effectivetemp}

We close this section with a brief discussion of the effective comoving temperature $T_0$  and the $\nu_H$ background energy density $\bar{\rho}_{\nu H}$ after equilibration, both of which appear in the effective collision integral~\eqref{eq:effectiveEoM}.

The $\nu_H$ background energy density appears in the energy density ratio $(a^4 \bar{\rho}_{\nu H})/(a^4 \bar{\rho}_{\nu\phi})$, where we remind the reader that $\bar{\rho}_{\nu\phi}$ stands for the total energy density in the combined $(\nu_H, \nu_l, \phi)$-system and is, leaving aside dilution due to expansion, equal to the pre-decay energy density in $\nu_H$ and $\nu_l$ .  This energy density ratio generally evolves with time, but at a rate much smaller than the Hubble expansion $H$ in the steady-state/quasi-equilibrium regime of our assumptions; Ref.~\cite{Barenboim:2020vrr} finds numerically a fairly constant 
\begin{equation}
\frac{(a^4 \bar{\rho}_{\nu H})}{(a^4 \bar{\rho}_{\nu\phi})} \sim \frac{1}{3} \to \frac{1}{2}
\end{equation}
in the time frame of interest (i.e., where $\Gamma_{\rm dec} \gg H$ and the ultra-relativistic approximation is well satisfied).  In our analysis, we take this ratio to be $1/3$.

The temperature~$T_0$ is the effective comoving temperature of the combined $(\nu_H, \nu_l, \phi)$-system, and should be lower than the pre-decay neutrino comoving temperature $T_{\nu,0} \simeq 1.95~{\rm K}\simeq 1.68 \times 10^{-4}$~eV.  To estimate $T_0$,  we note that energy conservation imposes the relation 
$a_{\rm pre}^4 \left(\bar{\rho}_{\nu H, {\rm pre}} + \bar{\rho}_{\nu l, {\rm pre}} \right) = a_{\rm post}^4 \left(\bar{\rho}_{\nu H, {\rm post}} + \bar{\rho}_{\nu l, {\rm post}} + \bar{\rho}_{\phi}\right)$
between the pre-decay and post-decay energy densities.
Assuming Maxwell-Boltzmann statistics (i.e., $\bar{\rho} =3 T_0 \bar{n}/a$), this leads to
\begin{equation}
\label{eq:totnu}
\frac{T_0}{T_{\nu,0}} =  \frac{a_{\rm pre}^3 \left(\bar{n}_{\nu H, {\rm pre}} + \bar{n}_{\nu l, {\rm pre}}\right)}{a_{\rm post}^3  \left(\bar{n}_{\nu H, {\rm post}} + \bar{n}_{\nu l, {\rm post}} + \bar{n}_{\phi} \right)}.
\end{equation}
The two-body decay also stipulates number conservation laws, namely, 
$a_{\rm pre}^3 \bar{n}_{\nu H,{\rm pre}}   = a_{\rm post}^3 \left(\bar{n}_{\nu H,{\rm post}} + \bar{n}_\phi\right)$ and 
$a_{\rm pre}^3 \bar{n}_{\nu l,{\rm pre}}  = a_{\rm post}^3 \left(\bar{n}_{\nu l,{\rm post}}  - \bar{n}_\phi\right)$, which, upon substitution into Eq.~\eqref{eq:totnu}, yields
\begin{equation}
\begin{aligned}
\frac{T_0}{T_{\nu,0} } & = 1 -  \frac{ \bar{n}_{\phi}}{\bar{n}_{\nu H, {\rm post}} + \bar{n}_{\nu l, {\rm post}} + \bar{n}_{\phi} } \\
& \simeq 1 -  \frac{ \bar{\rho}_{\phi}}{\bar{\rho}_{\nu\phi }}. 
\end{aligned}
\end{equation}
Numerically, Ref.~\cite{Barenboim:2020vrr} finds the energy density ratio to be at most $\bar{\rho}_{\phi}/\bar{\rho}_{\nu\phi } \sim 0.1$ in the steady-state regime of interest.  Therefore, in this analysis, we take $T_0$ to be  the same as the pre-decay neutrino comoving temperature~$T_{\nu,0}$.

\section{Statistical inference}
\label{sec:mcmc}

Having derived the effective equations of motion  in a way consistent with neutrino oscillation data, we wish now to put constraints on the neutrino rest-frame lifetime~$\tau_0$ using cosmological data.  

Before we describe our data analysis, observe that the effective collision integral~\eqref{eq:effectiveEoM} can be broken down in terms of its dependence on the scale factor~$a$ into
\begin{equation}
\left(	\frac{{\rm d} {\cal F}}{{\rm d} \tau} \right)_{C,\ell} = -  \alpha_\ell \, a^6  \, Y\,  \mathscr{F} \left(a X \right) {\cal F}_\ell.
\label{eq:effectivecollisionadep}
\end{equation}
Here,
\begin{equation}
X  \equiv \frac{m_{\nu H}}{T_0}  \simeq 298 \,  \left( \frac{m_{\nu H}}{0.05~{\rm eV}}\right)\,  \left( \frac{T_{\nu,0}}{T_0}\right)\label{eq:capX}
\end{equation}
is an effective mass parameter, while
\begin{equation}
	\begin{aligned}
Y & \equiv \,  \frac{C}{12} \, \frac{(a^4 \bar{\rho}_{\nu H})}{(a^4 \bar{\rho}_{\nu \phi})} \,  \Phi\left(\frac{m_{\nu l}}{m_{\nu H}}\right) \left( \frac{m_{\nu H}}{T_0}\right)^5  \Gamma^0_{\rm dec}  \\
& \simeq 6.55\, C \times 10^{10} \,  \left( \frac{(a^4 \bar{\rho}_{\nu H})}{(a^4 \bar{\rho}_{\nu \phi})} \Big/ \frac{1}{3}\right) \left( \frac{T_{\nu,0}}{T_0}\right)^5 \label{eq:capY}\\
& \hspace{23mm} \times \Phi\left(\frac{m_{\nu l}}{m_{\nu H}}\right) \,  \left( \frac{m_{\nu H}}{0.05~{\rm eV}}\right)^5\,   \Gamma^0_{\rm dec}
\end{aligned}
\end{equation}
is an effective transport rate, with $C=1$ for the one parent/one daughter scenario (``one  decay channel''), and $C=2$ if we wish to use the three-state to two-state approximation to describe the case of two parents/one daughter or one parent/two daughters (``two  decay channels'').  Both $X$  and $Y$ contain only quantities that are taken to be constant in time in our analysis (see Sect.~\ref{sec:effectivetemp}), and together these two effective parameters determine completely the phenomenology of the effective collision integral~\eqref{eq:effectiveEoM}.

\subsection{Model parameter space} 
\label{sec:modelspace}

To derive credible limits on the rest-frame lifetime~$\tau_0$, we consider two one-variable extensions to the standard six-variable \(\Lambda \textrm{CDM}\) fit, scenarios A and B, whose parameter spaces are spanned respectively by
\begin{equation}
\begin{aligned}
\label{eq:fitparamsA}
&{\fett \theta}_A  = \{\omega_{b}, \omega_{c}, \theta_s, \tau_{\rm reio}, n_s, \ln (10^{10} A_s), Y \}, \\
 & {\fett {\theta}_A}^{\rm fixed} = \{N_{\rm fs} =1, N_{\rm int} = 2.0440, X=X_{\rm scan}\},
 \end{aligned}
 \end{equation} 
 and
 \begin{equation}
 \begin{aligned}
 \label{eq:fitparamsB}
&\fett{\theta}_B = \{\omega_{b}, \omega_{c}, \theta_s, \tau_{\rm reio}, n_s, \ln (10^{10} A_s), Y \}, \\
& {\fett{\theta}_B}^{\rm fixed} = \{N_{\rm fs} =0, N_{\rm int} = 3.0440,X=X_{\rm scan}\}.
\end{aligned}
\end{equation}
In both scenarios, the variables are the physical baryon density parameter  \(\omega_{b}\), the physical cold dark matter density parameter  \(\omega_{c}\), the angular sound horizon $\theta_s$, the optical depth to reionisation $\tau_{\rm reio}$,  the spectral index of the primordial curvature power spectrum \(n_{s}\) and its amplitude \(A_s\) at the  pivot scale $k_{\rm piv} = 0.05~{\rm Mpc}^{-1}$, and the effective transport rate~$Y$.  The scenarios differ in the choices of fixed parameters in the neutrino sector.

\paragraph{Scenario~A} Here, the numbers of free-streaming and interacting  neutrinos are always fixed at $N_{\rm fs}=1$ and $N_{\rm int}=2.0440$ respectively (such that $N_{\rm fs}$ and $N_{\rm int}$ add up to the standard  $N_{\rm eff}^{\rm SM} = 3.0440$~\cite{Froustey:2020mcq,Bennett:2020zkv}), while the effective mass parameter $X$ is fixed at a value $X_{\rm scan}$ to be discussed below.
We model all neutrinos as massless states in \CLASS{}, in keeping with the limits of validity of our effective equations of motion~\eqref{eq:effectiveEoM}. Planck 2018 CMB data alone currently constrain the neutrino mass sum to $\sum m_\nu \lesssim 0.3$~eV~(95\%C.I.) in a $\Lambda$CDM+$m_\nu$ 7-variable fit~\cite{Planck:2018vyg}; in combination with non-CMB data, the bound tightens to $\sum m_\nu \lesssim 0.12$~eV~(95\%C.I.)~\cite{Planck:2018vyg}.
Therefore, as long as we fit CMB data only and cap the parent neutrino mass at $m_{\nu H}\simeq 0.1$~eV in the interpretation of our fit outcomes, our modelling is internally consistent.  Note that this is a lower cap than the limit  $m_{\nu H}\lesssim  0.2$~eV established  in Sect.~\ref{sec:physicalsystem}, which, we remind the reader, stems from requiring that the neutrinos stay ultra-relativistic up to recombination.%
\footnote{We do not consider the case of one {\it massive} free-streaming and two massless interacting neutrinos, because given current sensitivities, any bound on the neutrino mass derived under these assumptions cannot have an internally consistent interpretation that is also compatible with the mass spectra established by oscillations experiments.  To see this, suppose for instance the fit returns an upper  limit of $0.3$~eV on the mass of the free-streaming neutrino. Then,  by oscillation measurements the other two (interacting) neutrinos must also have masses in the $0.3$~eV ball park, which immediately conflicts with the $m_{\nu H} \lesssim 0.2$~eV limit of validity of our simplified modelling.   Furthermore, adding up the three masses gives a neutrino mass sum of $0.9$~eV---not the $0.3$~eV indicated by the modelling and hence the fit---and this is unlikely to be compatible with CMB data. In other words, the $0.3$~eV mass limit we started with in this example is actually meaningless.}

The mass cap also sets an upper limit on the value of $X_{\rm scan}$ that can be unambiguously explored with our modelling.  Specifically,  for the choice of $T_0 = T_{\nu,0}$, we are limited to the parameter range  
\begin{equation}
X_{\rm scan} \in [52,595],
\end{equation}
where the lower end corresponds to the smallest possible mass gap consistent with oscillation measurements, i.e., $\sqrt{\Delta m_{21}^2}=0.0087$~eV, and the upper limit is the mass cap discussed immediately above.  Then, with appropriate post-processing,  scenario A can be interpreted to represent six physically distinct cases,  summarised in Table~\ref{tab:cases}.

\begin{table}[t]
	\begin{center}
	\begin{tabular}{|c|c|cc|c|c|}
		\toprule
		   &  & FS & Decay & Gap& Min $m_{\nu H}^2$\\
		\toprule
		\multicolumn{6}{|c|}{\rotatebox[origin=c]{0}{\bf Scenario A: one decay channel}} \\
		\toprule
		\parbox[t]{4mm}{\multirow{4}{*}{\rotatebox[origin=c]{0}{A1}}} 
	  &  \parbox[t]{5mm}{\multirow{2}{*}{\rotatebox[origin=c]{0}{NO}}}   & $\nu_1$ & $\nu_3 \to \nu_2$ & $\Delta m_{32}^2|_{\rm N}$ &  \parbox[t]{14mm}{\multirow{2}{*}{\rotatebox[origin=c]{0}{$|\Delta m_{31}^2|_{\rm N}$ }}}  
	 	  \\
		\cline{3-5}
		     &  & $\nu_2$ & $\nu_3\to \nu_1$& $|\Delta m_{31}^2|_{\rm N}$ &    \\
		\cline{2-6}
		 &   \parbox[t]{5mm}{\multirow{2}{*}{\rotatebox[origin=c]{0}{IO}}}   & $\nu_2$ & $\nu_1 \to \nu_3$&$|\Delta m_{31}^2|_{\rm I}$&  $|\Delta m_{31}^2|_{\rm I}$   \\
				\cline{3-6}
		 &  & $\nu_1$ & $\nu_2 \to \nu_3$& $\Delta m_{23}^2|_{\rm I}$&  $\Delta m_{23}^2|_{\rm I}$   \\
		\cline{1-6}
		 A2
  &
		\parbox[t]{5mm}{\multirow{1}{*}{\rotatebox[origin=c]{0}{NO}}}  & $\nu_3$ & $\nu_2 \to \nu_1$ &   
		 \parbox[t]{10mm}{\multirow{2}{*}{\rotatebox[origin=c]{0}{$\Delta m_{21}^2$}}}  & $\Delta m_{21}^2$   \\
		\cline{1-4}
		\cline{6-6}
		A3 & \parbox[t]{5mm}{\multirow{1}{*}{\rotatebox[origin=c]{0}{IO}}}   & $\nu_3$ & $\nu_2 \to \nu_1$& &  $\Delta m_{23}^2|_{\rm I}$  \\
		\toprule
			\multicolumn{6}{|c|}{\rotatebox[origin=c]{0}{\bf Scenario B: two decay channels}} \\
		\toprule
		B1	
		 & \parbox[t]{5mm}{\multirow{1}{*}{\rotatebox[origin=c]{0}{NO}}}   & -- & $\nu_3 \to \nu_2,\nu_1$& 
		$|\Delta m_{31}^2|_{\rm N}$ &  $|\Delta m_{31}^2|_{\rm N}$  	\\
		 \cline{1-6}
		  B2 & \parbox[t]{5mm}{\multirow{1}{*}{\rotatebox[origin=c]{0}{IO}}}   & -- & $\nu_1,\nu_2 \to \nu_3$ & 	$|\Delta m_{31}^2|_{\rm I}$ & $|\Delta m_{31}^2|_{\rm I}$  	\\
		\toprule
	\end{tabular}
\end{center}
\caption{ All decay and free-streaming (FS) scenarios considered in this work, and their corresponding mass ordering (NO or IO),  decay mass gap, and minimum allowed parent neutrino mass $m_{\nu H}$.  We use the convention  $m_3 > m_2 > m_1$  for 
normal mass ordering  and $m_2 > m_1 > m_3$ for  inverted mass ordering, and  a common value $\Delta m_{21}^2 = 7.50 \times 10^{-5}$.  Note that  numerically we do not distinguish between  $|\Delta m_{31}^2|_{\rm N}$, $|\Delta m_{31}^2|_{\rm I}$,  $\Delta m_{32}^2|_{\rm N} \equiv |\Delta m_{31}^2|_{\rm N} -  \Delta m_{21}^2$, and $|\Delta m_{23}^2|_{\rm I} \equiv |\Delta m_{31}^2|_{\rm I} + \Delta m_{21}^2$: the difference between them is less than 5\%~\cite{deSalas:2020pgw,Esteban:2020cvm,Capozzi:2017ipn}.   This allows us to lump four physically distinct cases into one single scenario A1 and to use a single representative mass splitting $|\Delta m_{31}^2| = 2.5 \times 10^{-3}~{\rm eV}^2$.
	  The same is in principle also true for two physical cases of scenario B. Nonetheless, we distinguish between B1 and B2, because the former comprises two decay modes, while the latter has only one: this makes the definitions of the particle lifetime differ by a factor of two between the two cases.} \label{tab:cases}
\end{table}


\paragraph{Scenario~B}  This scenario has $N_{\rm int}=3.0440$ interacting neutrinos in the three-state to two-state approximation and no free-streaming neutrinos.  It represents two physically distinct cases, as summarised in Table~\ref{tab:cases}.  
Again, as with scenario A, to consistently interpret the fit outcomes requires that we limit the parent neutrino masses to $m_{\nu H}\lesssim 0.1$~eV.   This in turn limits the range of effective mass~$X$ to 
\begin{equation}
X_{\rm scan} \in [298,595]
\end{equation}
for the choice of $T_0 =T_{\nu,0}$, where the lower bound corresponds to the  mass gap  $\sqrt{|\Delta m_{31}^2|} = 0.05$~eV.  

\begin{figure}[t]
\begin{center}
\includegraphics[width=0.48\textwidth]{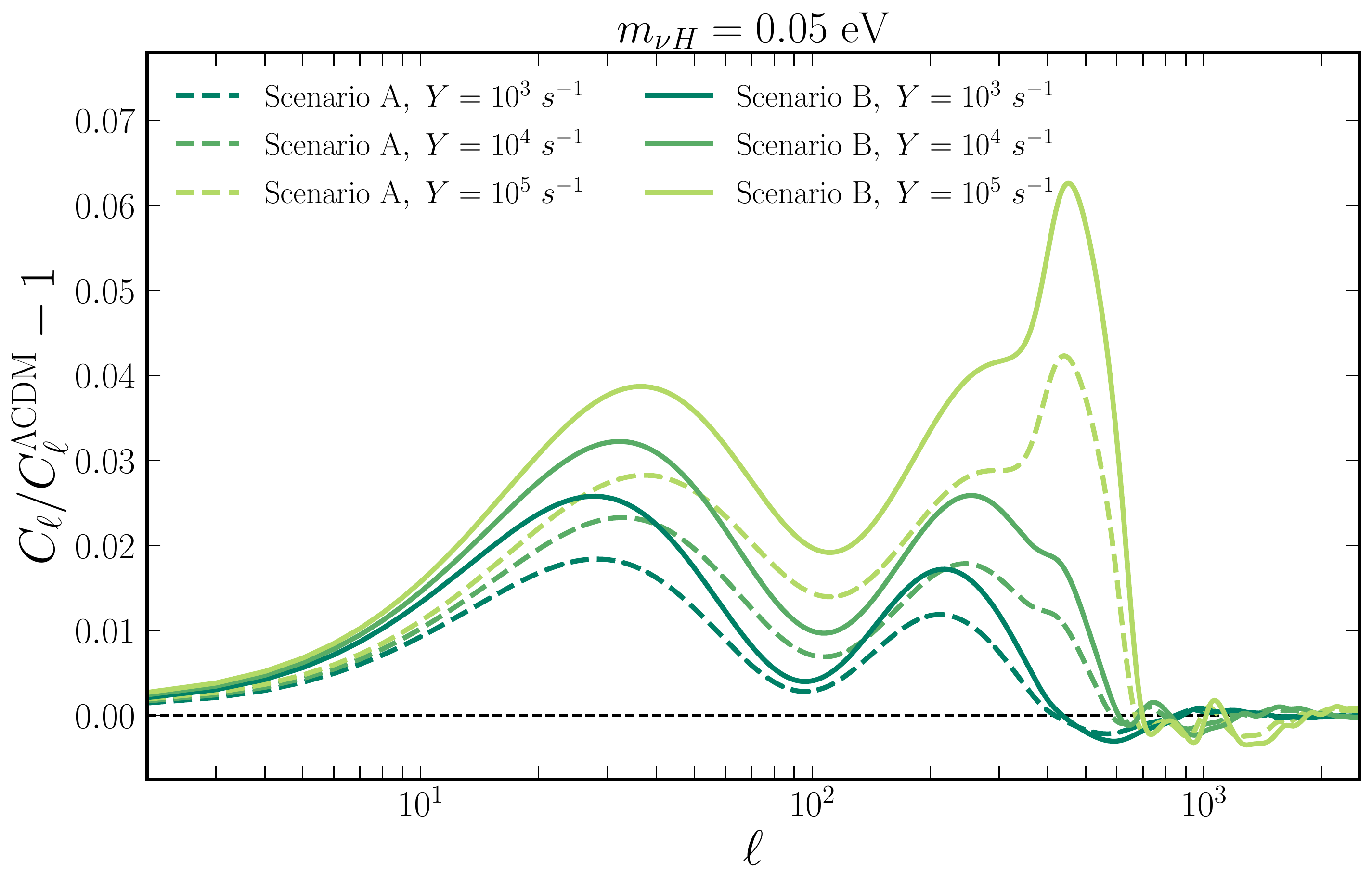}
\includegraphics[width=0.48\textwidth]{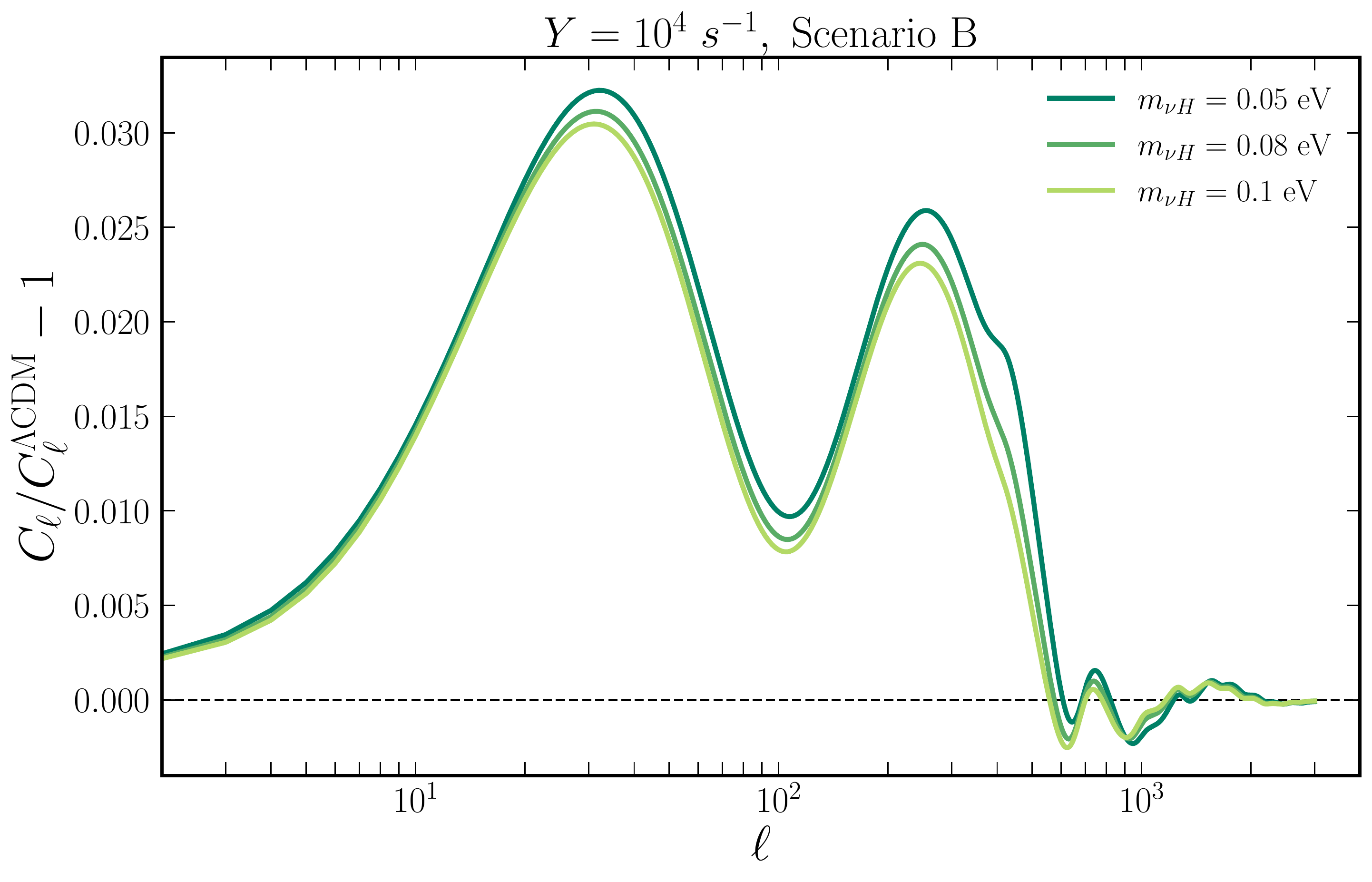}
\caption{CMB TT power spectra of ultra-relativistic invisible neutrino decay scenarios relative to the Planck 2018 best-fit $\Lambda$CDM cosmology. In each case, we vary only the effective mass parameter $X$ [Eq.~\eqref{eq:capX}] and effective transport rate $Y$ [Eq.~\eqref{eq:capY}], while keeping all other cosmological parameters fixed at their best-fit $\Lambda$CDM values.
\textit{Top}: Variations between scenarios A (solid line) and B (dashed line) for a range of $Y$ values, with $X$ fixed at $X = 298$ or, equivalently, $m_{\nu H}=0.05$~eV assuming $T_0 = T_{\nu,0}$. \textit{Bottom}:  Variations in scenario B with respect to the decaying neutrino mass $m_{\nu H}$, with $T_0=T_{\nu,0}$ and at a fixed the effective transport rate  $Y=10^4$~s$^{-1}$.\label{fig:TT}}
\end{center}
\end{figure}

 Figure~\ref{fig:TT} shows a selection of CMB $TT$ power spectra computed using our implementation of the effective equations of motion~\eqref{eq:effectivecollisionadep} in \CLASS{}~\cite{Blas:2011rf},%
 \footnote{Our modified version of \CLASS{} is available for download at \href{https://github.com/gpierobon/Class_NuDecay}{https://github.com/gpierobon/Class\_NuDecay}.}
 relative to the prediction for a reference $\Lambda$CDM cosmology.
For a fixed effective mass parameter $X = 298$ or, equivalently, $m_{\nu H}=0.05$~eV assuming $T_0 = T_{\nu,0}$, the top panel contrasts the signatures of ultra-relativistic neutrino decay for a range of effective transport rates~$Y$ in both scenarios~A and~B.  Observe that for the same value of $Y$, the signature of decay in scenario A is approximately a factor of $2/3$ times that of scenario B.  This is consistent with expectation, as only two out of three neutrinos interact in the former scenario, while all three neutrinos participate in the interaction in the latter case.  We can therefore expect constraints on $Y$ to be stronger in scenario B than in scenario A. 

The bottom panel of Fig.~\ref{fig:TT} shows variations in the decay signature with respect to the effective mass parameter~$X$, or $m_{\nu H}$ at a fixed $T_0$, for a fixed effective transport rate $Y = 10^4~{\rm s}^{-1}$ in scenario B.  Here, we see that the signature is weaker for larger values of $X$.  This is because $X$ appears only in the prefactor~$\mathscr{F}(a X)$ in the effective equation of motion~\eqref{eq:effectivecollisionadep}.  Since $\mathscr{F}(a X)$ is a decreasing function of its argument, for a fixed evolution history~$a(t)$, larger values of $X$ will generally lead to smaller prefactors and hence weaker signatures in the CMB anisotropies.
This also means that we can expect constraints on $Y$ to weaken with increasing $X$.


\subsection{Data and analysis}
\label{sec:data}

We compute the CMB temperature and polarisation anisotropies for a large sample of parameter values in the parameter spaces defined in Eqs.~\eqref{eq:fitparamsA} and \eqref{eq:fitparamsB} using our modified version of \CLASS{}. We test these outputs against the Planck 2018 data~\cite{Planck:2018vyg} using
\begin{enumerate}
	\item the TTTEEE likelihood at \(\ell \geq 30\), 
	\item the Planck low-\(\ell\) temperature+polarisation likelihood, and 
	\item the Planck lensing likelihood,
\end{enumerate}
a combination labelled ``TTTEEE+lowE+lensing'' in Ref.~\cite{Planck:2018vyg}. 
The parameter spaces of Eqs.~(\ref{eq:fitparamsA}) and (\ref{eq:fitparamsB}) are sampled as Monte Carlo Markov Chains (MCMC) generated with the package {\sc MontePython-3}~\cite{Brinckmann:2018cvx}.  We construct credible intervals using  {\sc GetDist}~\cite{Lewis:2019xzd}.

The prior probability densities employed in the analysis are always flat and linear in each parameter directions, with prior boundaries  as follows:
\begin{itemize}
	\item  For the variables $\omega_{b}, \omega_{c}, \theta_s, n_s$, and $\ln (10^{10}A_s)$, the priors are unbounded at both ends.
	\item  For the optical depth to reionisation, we impose  $\tau_{\rm reio} \in [0.01, \infty)$.  
	\item  For the effective transport rate, we use $Y \in [0,\infty)$ in scenario A and $Y \in [0,4000]$ in scenario B.
\end{itemize}
 We generate in excess of $250,000$ samples in each case (and over a million samples in some cases).  Convergence of the chains is defined by the Gelman-Rubin convergence criterion $R-1 < 0.015$.



\section{Results and discussions}
\label{sec:results}

\begin{figure}[t]
\begin{center}
\includegraphics[width=0.4\textwidth]{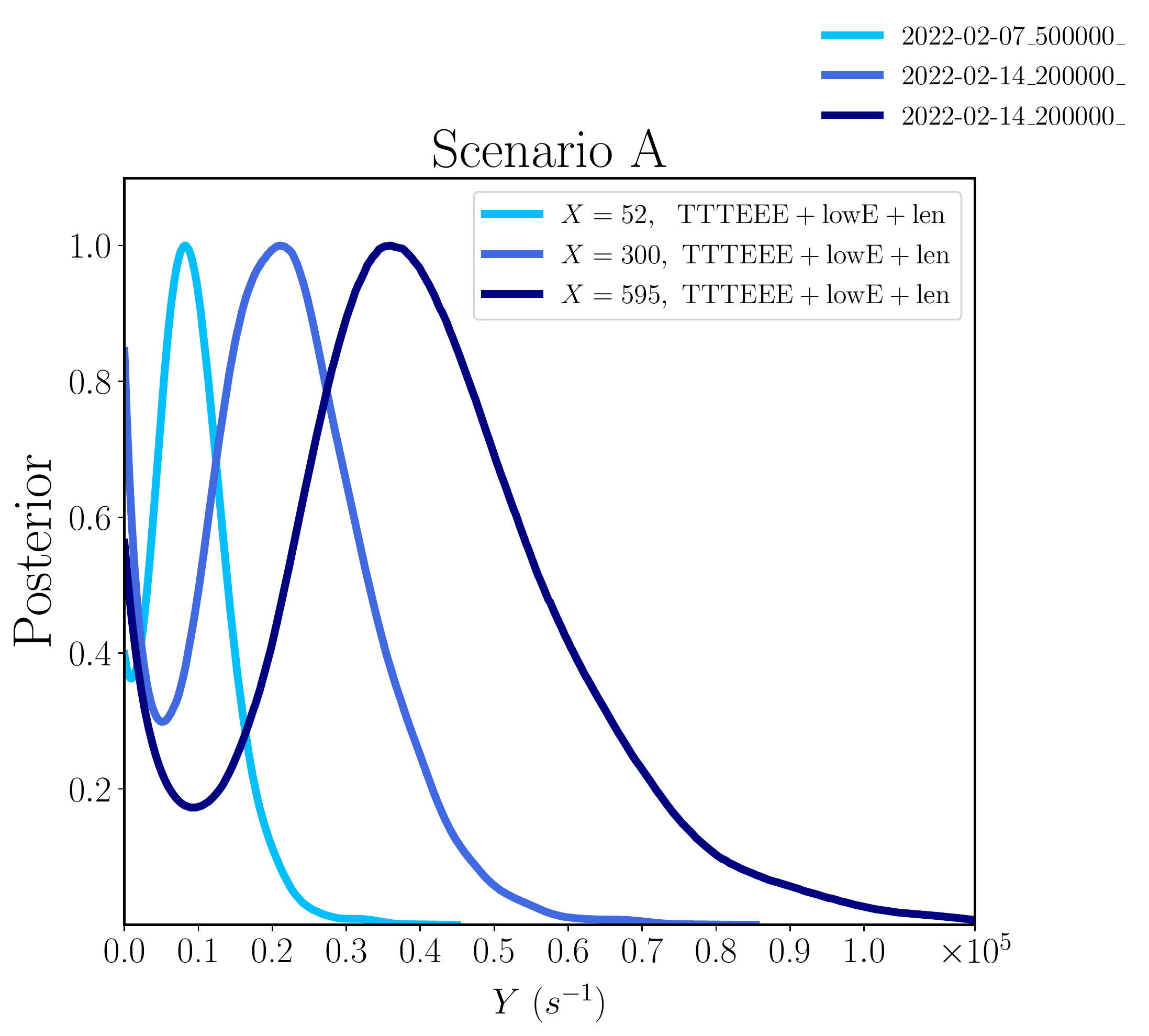}
\includegraphics[width=0.4\textwidth]{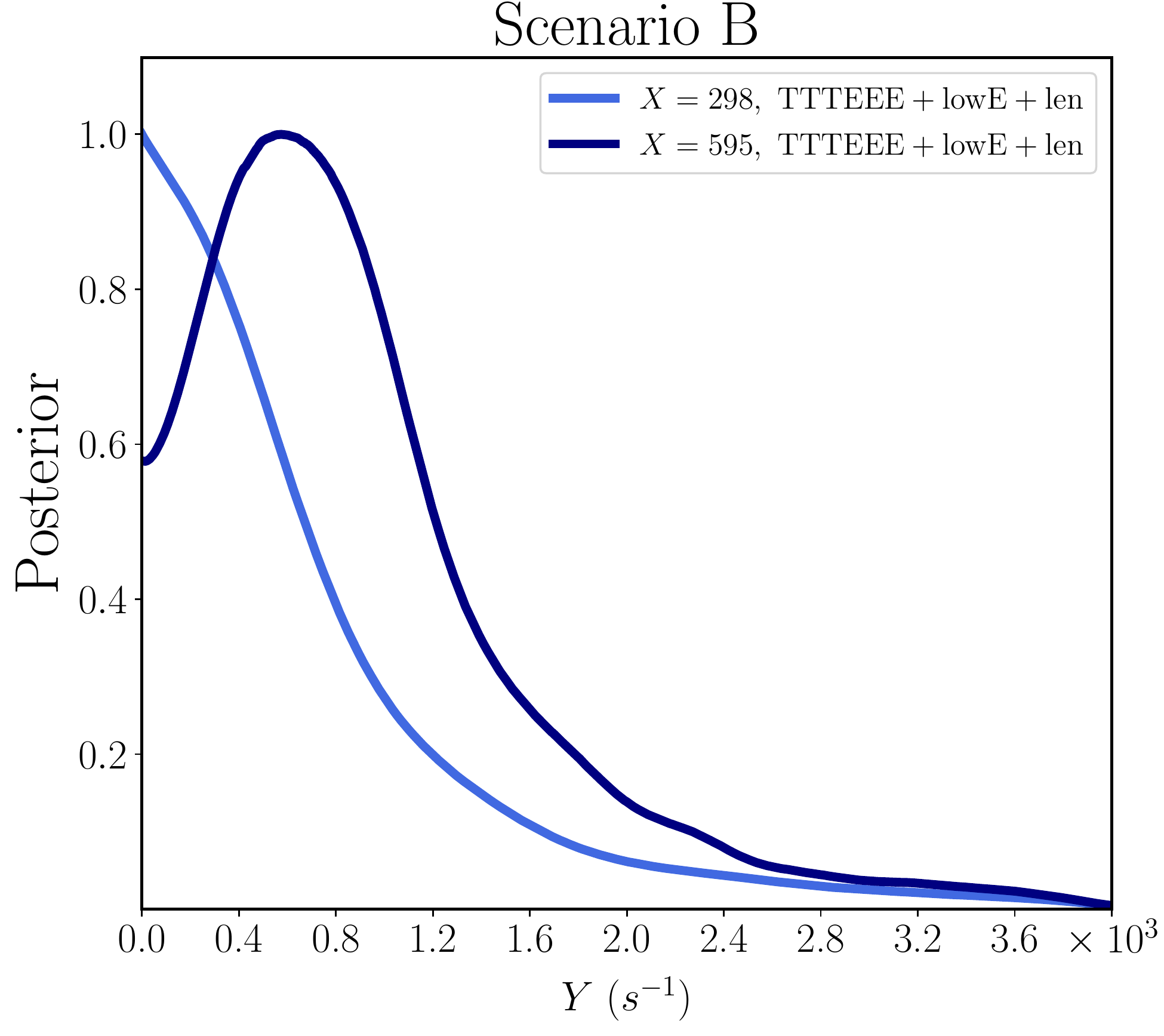}
\caption{1D marginalised posteriors for the effective transport rate~$Y$ [Eq.~\eqref{eq:capY}] obtained from our MCMC analysis for a range of fixed effective mass values $X=X_{\rm scan}$ in scenarios A (top panel) and B (bottom panel), whose parameter spaces are defined in Eqs.~\eqref{eq:fitparamsA} and~\eqref{eq:fitparamsB} respectively. The data used are the TTTEEE+lowE+lensing combination from the Planck 2018 data release~\cite{Planck:2018vyg}.~\label{fig:1Dposteriors}}
\end{center}
\end{figure}

Figure~\ref{fig:1Dposteriors} shows the 1D marginal posteriors for the effective transport rate~$Y$ for a range of fixed effective mass parameter values $X=X_{\rm scan}$ in both scenarios A (top panel) and B (bottom panel).   The corresponding one-sided 95\%  credible limits on $Y$ and the equivalent free-streaming redshift~$z_{\rm fs}$ in each case are summarised in Table~\ref{tab:Yconstraints}.  This redshift is defined via the condition 
\begin{equation}
\label{eq:zfs}
\Gamma_{\rm T}(z_{\rm fs})=H(z_{\rm fs}),
\end{equation}
where we have identified the transport rate $\Gamma_{\rm T}$ with the $\ell =2$ damping rate, i.e.,
\begin{equation}
\label{eq:gammaTT}
\Gamma_{\rm T} \equiv a^5 Y \mathscr{F}(a X).
\end{equation}
Physically, $z_{\rm fs}$ characterises the time at which the combined $(\nu_H,\nu_l,\phi)$-system switches from a free-streaming gas to a tightly-coupled fluid~\cite{Basboll:2008fx}.

\begin{table}[t]
	\begin{center}
	\begin{tabular}{|c|cc|c|c|}
		\toprule
		 & \multicolumn{2}{c|}{$X$} &  $Y$ (s$^{-1}$) & $z_{\rm fs}$\\
		  \cline{2-5}
	& Fixed & $m_{\nu H}$ (eV)	  & 95\% C.L. & 95\% C.L. \\
		\toprule
		\parbox[t]{2mm}{\multirow{3}{*}{\rotatebox[origin=c]{90}{\bf A}}} & $52$ & $0.0087$  & $< 18611.32$ & $<2457$ \\
		\cline{2-5}
		& $300$ & $0.05$  & $< 41490.24$ & $<2450$ \\
		\cline{2-5}
		&595& 0.1  & $< 76221.34$ & $<2476$\\
		\toprule
				\parbox[t]{2mm}{\multirow{2}{*}{\rotatebox[origin=c]{90}{\bf B}}} & 298 & 0.05   & $< 2286.08$ & $<1509$\\	
			\cline{2-5}
		&595 &0.1  & $< 2287.58$ & $<1338$ \\
		\toprule
	\end{tabular}
\end{center}
\caption{Marginalised one-sided 95\% credible limits on the effective transport rate~$Y$ and the equivalent free-streaming redshift $z_{\rm fs}$ [Eq.~\eqref{eq:zfs}] for fixed values of the effective mass parameter $X$ (and their equivalent parent neutrino mass $m_{\nu H}$ assuming $T_0=T_{\nu,0}$) in scenarios A and B.\label{tab:Yconstraints}}
\end{table}

Firstly, we observe that within each scenario, the constraint on $Y$ weakens with increasing $X$ and that for the same $X$, the scenario B constraint is typically an order of magnitude tighter than its scenario A counterpart.  Both trends are consistent with expectations (see Sect.~\ref{sec:modelspace}). Likewise, the constraints on the free-streaming redshift~$z_{\rm fs}$ are independent of the choice of $X$ within each scenario---to better than 1\% in scenario A and about 10\% in scenario B---as one would expect from the definition~\eqref{eq:zfs}.

Having established these $z_{\rm fs}$ bounds also allows us to roughly map our bounds on $Y$ in Table~\ref{tab:Yconstraints} to a simple expression
\begin{equation}
\label{eq:ylimitA}
Y \lesssim 5.3 \times 10^4 /\mathscr{F}(X/2460)~{\rm s}^{-1}  \end{equation}
in scenario A, and 
\begin{equation}
\label{eq:ylimitB}
Y \lesssim 2 \times 10^3 /\mathscr{F}(X/1500)~{\rm s}^{-1} 
\end{equation}
in scenario B.
In the case of the latter, our new constraint $z_{\rm fs} \lesssim 1500$ is clearly tighter than the previous limit $z_{\rm fs} \lesssim 1965$~\cite{Archidiacono:2013dua}. The difference is likely attributable to improved precision of the Planck 2018 CMB data over the 2013 data used in Ref.~\cite{Archidiacono:2013dua}, although of course the different time dependences of the transport rates 
assumed in the analyses---$\propto a^3$ in~\cite{Archidiacono:2013dua} vs our $\propto a^5 \mathscr{F}(a X)$ in Eq.~\eqref{eq:gammaTT}---as well as the assumption of instantaneous recoupling in Ref.~\cite{Archidiacono:2013dua} could also have played a role.

Secondly, while we have quoted in Table~\ref{tab:Yconstraints} one-sided 95\% credible limits on $Y$ and $z_{\rm fs}$, it is evident in Fig.~\ref{fig:1Dposteriors} that in almost all cases the 1D posterior for $Y$ peaks at a nonzero value.  Indeed, Ref.~\cite{Escudero:2019gfk} also found a similar feature, despite using a transport rate with a $\propto a^3$ time dependence [as opposed to our $\propto a^5 \mathscr{F}(a X)$ rate]. Reference~\cite{Archidiacono:2013dua} likewise found a shifted peak in their analysis of the Planck 2013 data for $z_{\rm fs}>0$, notwithstanding the assumption of instantaneous recoupling.  

However, in all cases the peak shift of either $Y$ or $z_{\rm fs}$ from zero is  statistically insignificant (i.e., $<2\sigma$) and  appears to be driven by CMB polarisation data~\cite{Escudero:2019gfk}.  We therefore dwell no further on the subject, except to note that the shifted $Y$ peak is accompanied by a marginal increase in the inferred scalar spectral index~$n_s$---a trend also observed in Ref.~\cite{Escudero:2019gfk}---and consequently the small-scale RMS fluctuation $\sigma_8$, in comparison with their $Y=0$ counterparts derived from the same data combination.
There are otherwise no strong degeneracies between $Y$ and other base $\Lambda$CDM parameters.  The interested reader may consult \ref{sec:parameter} for more summary statistics of our parameter estimation analyses.  Here, we conclude in view of the degeneracy directions that adding baryon acoustics oscillations measurements to the analysis is unlikely to improve the bound on $Y$, while a low-redshift small-scale $\sigma_8$ measurement could conceivably tighten the bound.%
\footnote{Note that we have had to apply a prior on the effective transport parameter~$Y$ in scenario B of $Y \in [0,4000]$, as opposed to allowing $Y$ to be unbounded, i.e., $Y \in [0,\infty)$,  as in scenario A. This is because,  had we not applied the restrictive prior, the $Y$ parameter in the case of $X=298$ would have 
exhibited a strong but accidental degeneracy with the spectral index $n_s$ in the high-$Y$ tail.  The tail region is extensive, with $\chi^2$ values hovering around  $\Delta \chi^2 \simeq 20$
worse than the best-fit, but presents no meaningful second peak.  Had we used low-redshift $\sigma_8$ measurements this tail would likely have been cut off naturally.  This tail feature is not present in the $X=595$ case.  But we have nonetheless applied the same restrictive prior on $Y$ here for consistency within scenario B.  Without the prior the 1D marginalised 95\% bound on $Y$ in the $X=595$ case would relax from $Y \lesssim 2288$ to $Y \lesssim 2546$.
}


\subsection{Constraints on the neutrino lifetime}

\begin{figure*}[t]
    \centering
    \includegraphics[width=\textwidth]{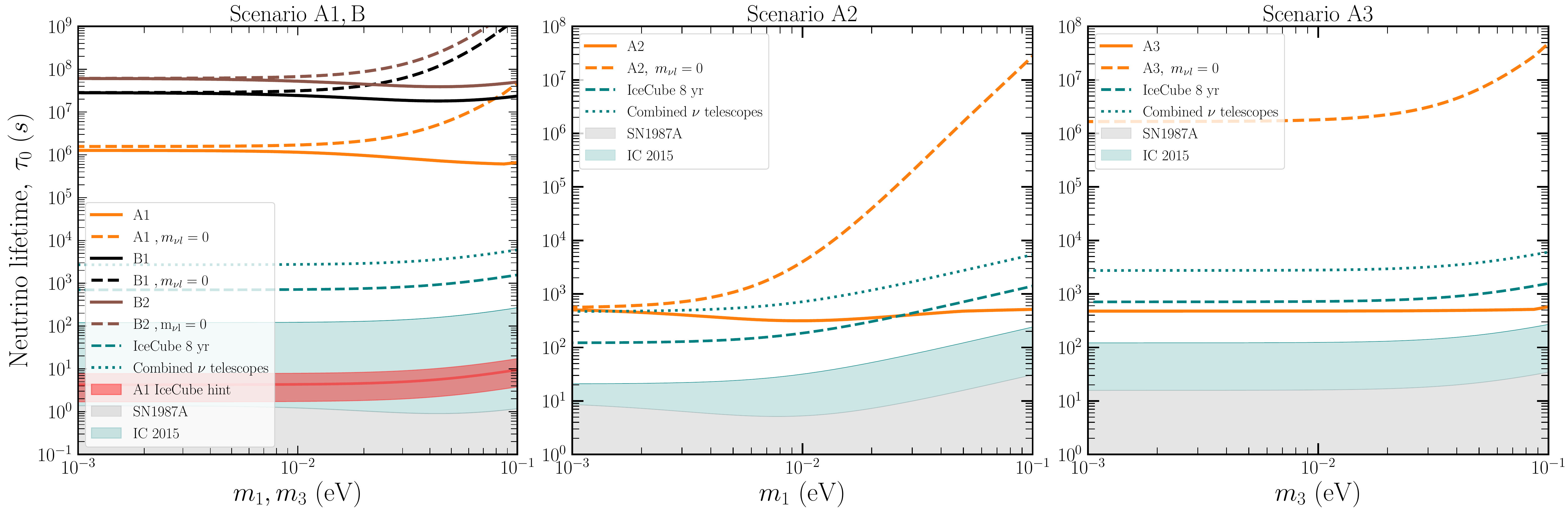} 
    \caption{Lower bounds on the neutrino lifetime converted from the 95\% credible limits on the effective transport rate $Y$ (Table~\ref{tab:Yconstraints}) for the three single-decay-channel cases of scenario A and the two two-decay-channel cases of scenario B (see Table~\ref{tab:cases}), as functions of the lightest neutrino mass $m_1$ (NO) or $m_3$ (IO).  Orange (black) solid lines denote the revised bounds of this work for scenarios A (B), while the dotted lines represent the na\"{\i}ve constraints we would have obtained had we assumed a massless daughter neutrino $m_{\nu l}=0$ in the phase space factor $\Phi(m_{\nu l}/m_{\nu H})$ [Eq.~\eqref{eq:phasespace}]. The interpretation of these bounds in conjunction with Table~\ref{tab:cases} is unambiguous in scenario~A.  In scenario B1 the bounds apply to $\nu_3$, while in B2 they apply to both $\nu_1$ and $\nu_2$ assuming equal lifetimes.
        We additionally show currents bounds from SN1987A \cite{Kachelriess:2000qc} and from the IceCube 2015 data analysed in Ref.~\cite{Song:2020nfh} (``IC 2015'') assuming the flavour composition of a full pion decay scenario, i.e., $(f_e,f_{\mu},f_{\tau})=(1,2,0)$, at the source,  as well as the projected reach of future IceCube runs (``IceCube 8 yr'') and a combination of future neutrino telescopes (including Baikal-GVD, KM3NeT, P-ONE,  TAMBO,  and  IceCube-Gen2), the latter of which also assumes improved oscillations parameter measurements from JUNO, DUNE, and Hyper-Kamiokande. See Ref.~\cite{Song:2020nfh} for details. Also displayed in red in the left panel is the invisible decay scenario proposed in Ref.\cite{Denton:2018aml} that would solve the tension between cascade and track data sets of IceCube.
    \label{fig:scenarioA}} 
\end{figure*}

To reinterpret our marginalised one-sided 95\% credible limits on the effective transport rate $Y$ (Table~\ref{tab:Yconstraints}) as limits on the rest-frame neutrino lifetime $\tau_0$, we first interpolate the constraints on~$Y$ as a function of the effective mass parameter $X$ (or, equivalently, the mass of the decaying neutrino~$m_{\nu H}$)  within each scenario.  Then, using Eq.~\eqref{eq:capY} these upper limits on $Y$ constraints can be converted into lower bounds on $\tau_0 = 1/\Gamma_{\rm dec}^0$ for any neutrino mass spectrum desired. Alternatively, we can use Eq.~\eqref{eq:capY} to directly translate the approximate limits of Eqs.~\eqref{eq:ylimitA} and~\eqref{eq:ylimitB} into 
\begin{equation}
\begin{aligned}
\label{eq:tau0limitA}
\tau_0 \gtrsim \,\, &  1.2 \times 10^6 \, \mathscr{F}
\left[0.12\left(\frac{m_{\nu H}}{0.05 {\rm eV}} \right)\right]\\
&  \hspace{23mm}   \times \Phi\left(\frac{m_{\nu l}}{m_{\nu H}}\right) \,  \left( \frac{m_{\nu H}}{0.05~{\rm eV}}\right)^5~{\rm s}
\end{aligned}
\end{equation}
for scenario A, and 
\begin{equation}
\begin{aligned}
\label{eq:tau0limitB2}
\tau_0 \gtrsim \,\, &  6.6 \times 10^7 \, \mathscr{F}
\left[0.2 \left(\frac{m_{\nu H}}{0.05 {\rm eV}} \right)\right]\\
&  \hspace{23mm}   \times \Phi\left(\frac{m_{\nu l}}{m_{\nu H}}\right) \,  \left( \frac{m_{\nu H}}{0.05~{\rm eV}}\right)^5~{\rm s}
\end{aligned}
\end{equation}
for scenario B2.

Note that in the case of B1, the lifetime of $\nu_3$ needs to be computed from $\tau_0 = 1/(2 \Gamma_{\rm dec}^0)$, because of its two distinct decay modes.  See Table~\ref{tab:cases}.

Figure~\ref{fig:scenarioA} shows the limits on $\tau_0$ constructed from interpolation for the physical cases listed in Table~\ref{tab:cases}, as functions of the lightest neutrino masses $m_1$ (NO) or $m_3$ (IO). Also displayed in the same plots are the na\"{\i}ve constraints on $\tau_0$ we would have obtained had we set the mass of the daughter neutrino to zero in the phase space factor $\Phi$.  For reference we also show current constraints from SN1987A~\cite{Kachelriess:2000qc} and IceCube~\cite{Song:2020nfh}, as well as projected constraints from future measurements by IceCube and other neutrino telescopes~\cite{Song:2020nfh}.%
\footnote{The present and projected constraints from IceCube and other neutrino telescopes shown in Fig.~\ref{fig:scenarioA} assume (i)~normal mass ordering, (ii)~two decaying neutrinos $\nu_2$ and $\nu_3$, and (iii) rest-frame lifetime-to-mass ratios satisfying $\tau_3/m_3=\tau_2/m_2$.  These assumptions imply a hierarchy of the associated transport rates, i.e., $\Gamma_{{\rm T},3 \to 1},\Gamma_{{\rm T},3 \to 2} \gg \Gamma_{{\rm T},2 \to 1}$, and are hence largely consistent with the modelling of the NO cases of our scenarios A1 (if only one of $\nu_3 \to \nu_1$ and $\nu_3 \to \nu_2$ is present) and~B1 (if both are present).  Less clear, however, is how to reconcile these assumptions with those behind our scenarios A2 and A3, as well as the IO cases of scenarios A1 and B2. Nonetheless, the decay modelling of Ref.~\cite{Song:2020nfh} suggests that switching the model assumption from two decaying and one stable neutrinos to one decaying and two stable neutrinos will not likely have a drastic  effect [i.e., no more than ${\cal O}(1)$] on the resulting  constraints on $\tau_0$.
We therefore opt to show the neutrino telescope constraints in the middle and right panels of Fig.~\ref{fig:scenarioA} despite the inconsistency, but caution the reader that these constraints must be taken as indicative only.\label{footnote:ice}}

Relative to our previous estimate of $\tau _0 \gtrsim  (4 \times 10^5 \to 4 \times 10^6 (m_{\nu H}/0.05 {\rm eV})^5$~s~\cite{Barenboim:2020vrr}, the new na\"{\i}ve (i.e., $m_{\nu l}=0$) constraints are comparable to the high end at $m_{\nu H} \simeq 0.05$~eV in scenario A (one decay channel) and a factor of 50 tighter  than the high end in scenario B (two decay channels) at the same mass point.  These differences can be attributed to (i)~different prefactors in the relation between the effective transport rate and the rest-frame decay rate [Eq.~\eqref{eq:capY}], and (ii)~the fact that the old estimate had been obtained from a rough conversion of the $z_{\rm fs}$ bound of Ref.~\cite{Archidiacono:2013dua}---itself derived from older CMB data---using the $\Gamma_{\rm T} (z_{\rm fs})=H(z_{\rm fs})$ condition.
Furthermore, because the constraints on $Y$ weaken with $m_{\nu}$ [due to  $\mathscr{F}(a m_\nu/T_0)$ in Eq.~\eqref{eq:effectivecollisionadep}], the scaling of the na\"{\i}ve $\tau_0$ bounds with $m_{\nu H}$ in fact follows more closely $\propto m_{\nu H}^4$ than $\propto m_{\nu H}^5$ in the mass range of interest (as can be seen most clearly in the middle panel of Fig.~\ref{fig:scenarioA}).

However, once a finite $m_{\nu l}$ has been accounted for via the new phase space factor $\Phi$ in accordance with oscillation measurements,  the lifetime bounds on the decaying neutrino relax significantly in all physical cases.  Specifically, if the parent-daughter neutrino mass gap corresponds to the atmospheric mass splitting (i.e., scenarios A1 and B), then irrespective of the neutrino mass ordering we see that the bound on $\tau_0$ at any one mass point weakens by up to a factor of 50 relative to the na\"{\i}ve bound for a lightest neutrino mass~$m_1$ (NO) or~$m_3$ (IO) not exceeding $0.1$~eV.  
    
Even more dramatic relaxations can be seen in those cases in which the mass gap is determined by the solar mass splitting, i.e., scenarios A2 and A3.  Here, the $\tau_0$ constraints weaken by four to five orders relative to the na\"{\i}ve bounds in the case of IO (i.e., scenario A3) across the whole $m_3$ range of interest.  In the case of NO (i.e., scenario A2), the relaxation also reaches five orders of magnitude, although for a smaller range of $m_1$.

Interestingly, inclusion of the new phase space factor~$\Phi$ also translates into revised lifetime constraints that are remarkably independent of the  neutrino mass.  We have seen previously in Sect.~\ref{sec:coll} that $\Phi$ asymptotes to $(1/3) (\Delta m_\nu^2/m_{\nu H}^2)^2$ in the limit $\Delta m_\nu^2 \equiv m_{\nu H}^2 - m_{\nu l}^2 \ll m_{\nu H}^2$.  Since the na\"{\i}ve bounds on $\tau_0$ scale effectively as $m_{\nu H}^4$ in the mass range of interest, we see immediately that including $\Phi$ must reduce the dependence of the final $\tau_0$ bounds to $\propto (\Delta m_\nu^2)^2$, which, depending on the neutrino pair participating in the decay, is always fixed by either the solar or the atmospheric squared neutrino mass splitting.   Indeed, we find in scenarios A2 and A3 a fairly mass-independent revised constraint of $\tau_0 \gtrsim (400 \to 500)$~s, while in scenario A1 the bound is some $(|\Delta m_{31}^2|/\Delta m_{21}^2)^2 \sim 10^3$ times tighter at $\tau_0 \gtrsim (6 \to 10)\times 10^5$~s.  The two scenario B bounds are tighter still at $\tau_0 \gtrsim (2 \to 6) \times 10^7$~s, but apply to two decay channels with a near-common atmospheric mass gap.

In the case of A2 and A3, we note while bearing in mind the caveats of Footnote~\ref{footnote:ice}
that our revised CMB constraints on $\tau_0$ are merely a factor of a few more stringent than current bounds derived in Ref.~\cite{Song:2020nfh} from the 2015 IceCube measurements of  the flavour composition of 
TeV--PeV-energy astrophysical neutrinos~\cite{IceCube:2015gsk}, assuming the flavour ratio of a full pion decay scenario, i.e.,  $(f_e,f_{\mu},f_{\tau})=(1,2,0)$, at the source.
Indeed, according to the projections of Ref.~\cite{Song:2020nfh}, the sensitivity to the neutrino lifetime of IceCube alone based on a combination 8 years of starting events and through-going track will likely already supersede our CMB constraints  entirely in scenario A3 and partially in scenario A2 in the mass range presented.  Together with improved measurements of the neutrino mixing parameters by JUNO~\cite{JUNO:2015zny}, DUNE~\cite{DUNE:2020lwj}, and  Hyper-Kamiokande~\cite{Hyper-Kamiokande:2018ofw}, even 
the A2 CMB constraints could become be entirely overtaken in the next two decades~\cite{Song:2020nfh} by data collected at an array of future neutrino telescopes such as Baikal-GVD~\cite{Baikal-GVD:2019fko}, KM3NeT~\cite{KM3Net:2016zxf}, P-ONE~\cite{P-ONE:2020ljt}, TAMBO~\cite{Romero-Wolf:2020pzh}, and IceCube-Gen2~\cite{IceCube-Gen2:2020qha}.  
In light of this possibility, it would be interesting to see if the neutrino telescope 
constraints on and projected sensitivities to $\tau_0$ would vary significantly if derived under model assumptions more consistent with our scenarios A2 and A3.
Likewise, it remains to be determined if future CMB measurements on a comparable timeline such as CMB-S4~\cite{Abazajian:2019eic} will be competitive in this space.

 Lastly, one may be tempted to use Eqs.~\eqref{eq:tau0limitA} and \eqref{eq:tau0limitB2} to extrapolate our lifetime limits to parent neutrino mass values beyond our scan range, i.e., $m_{\nu H} \gtrsim 0.1$~eV. Leaving aside that such large masses might run into problems with CMB  bounds, it is interesting to note the function $\mathscr{F}(x)$  naturally shuts down the anisotropy loss when the condition of ultra-relativistic decay cannot be satisfied, which translates in turn into a rapid deterioration of the lifetime bounds at large values of $m_{\nu H}$. In fact, for $m_{\nu H} \gtrsim 1$~eV in scenario A and $m_{\nu H} \gtrsim 0.5$~eV in scenario B, we expect the lifetime bounds to deteriorate as $\sim \exp(-x)/x^{0.5}$, where $x=x_A \simeq 0.12 \times (m_{\nu H}/0.05~{\rm eV})$ and $x=x_B \simeq 0.2 \times (m_{\nu H}/0.05~{\rm eV})$ for scenarios A and B respectively.  In other words, at such large $m_{\nu H}$ values, there are no neutrino lifetime constraints from free-streaming requirements.

\section{Conclusions}
\label{sec:conclusions}

We have considered in this work invisible neutrino decay $\nu_H \to \nu_l+\phi$ and its inverse process in the ultra-relativistic limit, and derived the effective Boltzmann hierarchy and associated $\ell$th anisotropy loss rate (or, equivalently, the transport rate~$\Gamma_{\rm T}$) for the coupled $(\nu_H,\nu_l,\phi)$-system relevant for CMB anisotropy computations.  Relative to our previous work~\cite{Barenboim:2020vrr} which assumed both decay products to be  massless, we have in this work allowed the daughter neutrino $\nu_l$ to remain massive. 

We find that a nonzero $m_{\nu l}$ introduces a new phase space suppression factor $\Phi(m_{\nu l}/m_{\nu H})$ [Eq.~\eqref{eq:phasespace}] in the transport rate, which in the limit of a small squared mass gap,  $\Delta m_\nu^2 \equiv m_{\nu H}^2-m_{\nu l}^2 \ll m_{\nu H}^2$, asymptotes to $(1/3) (\Delta m_\nu^2/m_{\nu H}^2)^2$.  Thus, the transport rate $\Gamma_{\rm T}$ is related to the rest-frame lifetime $\tau_0$ of the parent neutrino $\nu_H$ roughly as $\Gamma_{\rm T} \sim (\Delta m_\nu^2/m_{\nu H}^2)^2 (m_{\nu H}/E_\nu)^5(1/\tau_0)$, where the factor $(m_{\nu H}/E_\nu)^5$ was established previously in Ref.~\cite{Barenboim:2020vrr}.  Note that while we have derived our transport rates rigorously from first principles under a reasonable set of assumptions, it is possible to attribute the various components that make up the expressions to physical quantities such as the opening angle and the momenta of the decay products. We have dedicated \ref{sec:heuristic} to  interpreting our transport rate in terms of a random walk using these physical quantities, as well as explaining why the old random walk argument of Ref.~\cite{Hannestad:2005ex} is incomplete and hence yields a 
transport rate that is too large.
We emphasise that the new phase space factor $\Phi$ is {\it in addition} to the phase space suppression already inherent in the lifetime~$\tau_0$, and traces its origin to the momenta carried by the decay products in the decay rest frame.

Implementing our effective Boltzmann hierarchy~\eqref{eq:hierarchytotalF} and transport rates~\eqref{eq:effectivecollisionadep} into the CMB code \CLASS{}~\cite{Blas:2011rf}, we have performed a series of MCMC fits to the CMB temperature and polarisation anisotropy measurements by the Planck mission (2018 data release) and determined a new set of constraints on the neutrino lifetime~$\tau_0$ in a manner consistent with the neutrino mass splittings established by oscillation experiments. These new constraints are presented in  Fig.~\ref{fig:scenarioA} as functions of the lightest neutrino mass, i.e., $m_1$ in the normal mass ordering ($m_3> m_2>m_1$) and $m_3$ in the inverted mass ordering $(m_2 > m_1>m_3)$.  For a parent neutrino mass up to about $m_{\nu H} \simeq 0.1$~eV, we find that the presence of the new phase space factor weakens the CMB constraint on its lifetime by as much as a factor of 50 if the daughter neutrino mass $m_{\nu l}$ is separated from $m_{\nu H}$  by the atmospheric mass gap, and by up to five orders of magnitude if separated by the solar mass gap, in comparison with na\"{\i}ve limits that assume $m_{\nu l}=0$.

 The revised  constraints  are (i)~$\tau^0 \gtrsim (6 \to 10) \times 10^5$~s and $\tau^0 \gtrsim (400 \to 500)$~s if only one neutrino decays to a daughter neutrino separated by, respectively, the atmospheric and the solar mass gap, and  (ii)~$\tau^0 \gtrsim (2 \to 6) \times 10^7$~s  in the case of two decay channels with one near-common atmospheric mass gap.   Relative to existing constraints derived from other probes, these CMB limits are, despite significant revision, still globally the most stringent.  However, we note that the relaxation of the bounds has also opened up a swath of parameter space likely within the projected reach of  IceCube and other telescopes in the next two decades~\cite{Song:2020nfh}.    This may in turn call for closer scrutiny of the neutrino anisotropy loss modelling upon which
the derivation of the CMB constraints of this work is based. For example, where cosmological and astrophysical constraints meet, a more consistent treatment of neutrinos masses and quantum statistics may be required to accurately delineate the viable decay  parameter space.    We defer this investigation to a future work.

Finally, we remind the reader that while we have consistently accounted for the daughter neutrino mass according to experimentally measured mass splittings, we have assumed $m_{\phi}=0$ in the determination of the lifetime constraints.  This assumption need not be true, and a choice of $m_\phi$ satisfying $m_{\nu H} > m_\phi \gg m_{\nu l}$ in the regions (i)~$m_1, m_3 \lesssim 10^{-2}$~eV in scenarios A1 and B and (ii)~$m_1 \lesssim 2 \times 10^{-3}$~eV in A2 may in fact also relax the cosmological constraints on $\tau_0$ 
presented in Fig.~\ref{fig:scenarioA}. (Phase space considerations limit the range of permissible $m_\phi$ values to $m_\phi < m_{\nu H}- m_{\nu l}$ and hence their effects on the $\tau_0$ constraints where a finite $m_{\nu l}$ already produces a significant suppression, except for a few finely-tuned values in the vicinity of $m_\phi = \Delta m_\nu$.)  Such a relaxation may be similarly modelled using the phase space  factor $\Phi$ [Eq.~\eqref{eq:phasespace}] with $m_{\nu l} \to m_{\phi}$, although the resulting limits on $\tau_0$ would depend strongly on the choice of mass for the hypothetical $\phi$ particle.  We leave the exploration of $m_\phi$-dependent neutrino lifetime constraints  as an exercise to the interested reader.


\begin{acknowledgement}

We thank Jan Hamann for useful discussions on random walks.
JZC acknowledges support from an Australian Government Research Training Program Scholarship. IMO is supported by Fonds de la recherche scientifique (FRS-FNRS). GP is supported by a  UNSW University International Postgraduate Award.   Y$^3$W is supported in part by the Australian Government through the Australian Research Council's  Future Fellowship (project FT180100031).  This research is enabled by the Australian Research Council's Discovery Project (project DP170102382) funding scheme, and includes computations 
using the computational cluster Katana supported by Research Technology Services at UNSW Sydney.
\end{acknowledgement}

\bibliographystyle{utcaps}
\bibliography{Literature}


\appendix 
\onecolumn

\section{Collision integrals}
\label{sec:individualcollisionintegral}

The individual linear-order decay/inverse decay collision integrals for the $(\nu_H, \nu_l, \phi)$-system
have been derived in Ref.~\cite{Barenboim:2020vrr} assuming a Yukawa interaction.  Here, we present a generalised version of the result valid for any interaction structure, recast following the arguments of Sect.~\ref{sec:physicalsystem} such that the rest-frame decay rate $\Gamma_{\rm dec}^0$ (rather than the squared matrix element $|{\cal M}|^2$) appears in the integrals as a fundamental parameter.

Decomposed in terms of a Legendre series, the $\ell$th moments of the three collision integrals are given respectively by
\begin{equation}
	\begin{aligned}
		\left( \frac{{\rm d} f_{\nu H}}{{\rm d} \tau}\right)_{C,\ell}^{(1)} 
		(|\mathbf{q}_1|)
		=& \, \frac{2 a^2 m_{\nu H}^3 \Gamma^0_{\rm dec}}{\Delta (m_{\nu H},m_{\nu l},m_\phi)\epsilon_1 |\mathbf{q}_1|} \left \{-F_{\nu H,\ell}(|\mathbf{q}_1|) \int_{q_{2-}^{(\nu H)}}^{q_{2+}^{(\nu H)}} \mathrm{d}|\mathbf{q}_2| \, \frac{|\mathbf{q}_2|}{\epsilon_2} \,\Omega_1 \right. \\
		& \hspace{-4mm} \left.+  \int^{q_{2+}^{(\nu H)}}_{q_{2-}^{(\nu H)}} \mathrm{d} |\mathbf{q}_2|  \frac{|\mathbf{q}_2|}{\epsilon_2} \, P_\ell (\cos \alpha^*) \,   F_{\nu l,\ell}(|\mathbf{q}_2|) \, \Omega_2 
		+\int^{q_{3+}^{(\nu H)}}_{q_{3-}^{(\nu H)}} \mathrm{d} |\mathbf{q}_3|  \,  \frac{|\mathbf{q}_3|}{\epsilon_3} \,  P_\ell(\cos \beta^*) \,  F_{\phi,\ell}(|\mathbf{q}_3|) \, \Omega_3 \right\},
		\label{eq:C1_nuH}
	\end{aligned}
\end{equation}
\begin{equation}
	\begin{aligned}
		\left( \frac{{\rm d} f_{\nu l}}{{\rm d} \tau}\right)_{C,\ell}^{(1)} 
		(|\mathbf{q}_2|)
		=&\,
		\frac{2 a^2 m_{\nu H}^3  \Gamma^0_{\rm dec}}{\Delta(m_{\nu H},m_{\nu l},m_\phi) \epsilon_2 |\mathbf{q}_2|} \left\{
		\int_{q_{1-}^{(\nu l)}}^{q_{1+}^{(\nu l)}}  \mathrm{d} |\mathbf{q}_1| \,   \frac{|\mathbf{q}_1|}{\epsilon_1} \,  P_\ell(\cos \alpha^*) F_{\nu H,\ell}(|\mathbf{q}_1|) \, \Omega_1 \right.\\
		&\hspace{0mm}\left.-  F_{\nu l,\ell}(|\mathbf{q}_2|)
		\int_{q_{1-}^{(\nu l)}}^{q_{1+}^{(\nu l)}} \mathrm{d}|\mathbf{q}_1| \, \frac{|\mathbf{q}_1|}{\epsilon_1} \, \Omega_2  -  \int_{q_{3-}^{(\nu l)}}^{q_{3+}^{(\nu l)}}  \mathrm{d} |\mathbf{q}_3|  \,   \frac{|\mathbf{q}_3|}{\epsilon_3} \, P_\ell(\cos \gamma^*) F_{\phi,\ell}(|\mathbf{q}_3|)\, \Omega_3 \right\},
		\label{eq:C1_nul}
	\end{aligned}
\end{equation}
\begin{equation}
	\begin{aligned}
		\left( \frac{{\rm d} f_{\phi}}{{\rm d} \tau}\right)_{C,\ell}^{(1)} 
		(|\mathbf{q}_3|)
		= &\,\frac{4 a^2 m_{\nu H}^3 \Gamma^0_{\rm dec} }{\Delta(m_{\nu H},m_{\nu l},m_\phi) \epsilon_3 |\mathbf{q}_3|} \left\{
		\int_{q_{1-}^{(\phi)}}^{q_{1+}^{(\phi)}} \mathrm{d} |\mathbf{q}_1|  \,  \frac{|\mathbf{q}_1|}{\epsilon_1}  \, P_\ell(\cos \beta^*) F_{\nu H,\ell}(|\mathbf{q}_1|) \, \Omega_1 \right.\\
		& \hspace{0mm} \left.- \int_{q_{2-}^{(\phi)}}^{q_{2+}^{(\phi)}} \mathrm{d} |\mathbf{q}_2| \,   \frac{|\mathbf{q}_2|}{\epsilon_2} \, P_\ell(\cos \gamma^*)
		F_{\nu l,\ell}(|\mathbf{q}_2|) \, \Omega_2 
		- F_{\phi,\ell} (|\mathbf{q}_3|) \int_{q_{1-}^{(\phi)}}^{q_{1+}^{(\phi)}} \mathrm{d}|\mathbf{q}_1| \, \frac{|\mathbf{q}_1|}{\epsilon_1} \, \Omega_3 \right\}.
		\label{eq:C1_phi}
	\end{aligned}
\end{equation}
Here, 
\begin{equation}
    \begin{aligned}
		\cos{\alpha^*} =& \, \frac{2\epsilon_1 \epsilon_2 - a^2 (m_{\nu H}^2 + m_{\nu l}^2 - m_{\phi}^2)}{2|\mathbf{q}_1||\mathbf{q}_2| },\\
		\cos{\beta^*} =& \, \frac{2\epsilon_1 \epsilon_3 - a^2 (m_{\nu H}^2 - m_{\nu l}^2 + m_{\phi}^2)}{2|\mathbf{q}_1||\mathbf{q}_3| }, \\
		\cos \gamma^*  =&\, \frac{2\epsilon_2 \epsilon_3 -a^2(m_{\nu H}^2 - m_{\nu l}^2 - m_{\phi}^2)}{2|\mathbf{q}_2||\mathbf{q}_3|}
	\end{aligned}
\end{equation}
originate purely from kinematics; the phase space factors including quantum statistics and their corresponding no-quantum-statistics approximations ($\slashed{\rm qs}$) are
\begin{equation}
	\begin{aligned}
		\Omega_1 (|\mathbf{q}_2|, |\mathbf{q}_3|)  & = 1  -  \bar{f}_{\nu l}(|\mathbf{q}_2|)  +  \bar{f}_{\phi}(|\mathbf{q}_3|) \underset{\slashed{\rm qs}}{\to} 1,\\
		\Omega_2(|\mathbf{q}_1|, |\mathbf{q}_3|)  & =  \bar{f}_{\nu H}(|\mathbf{q}_1|)  +  \bar{f}_{\phi}(|\mathbf{q}_3|) \underset{\slashed{\rm qs}}{\to} \bar{f}_{\phi}(|\mathbf{q}_3|) , \\
		\Omega_3(|\mathbf{q}_1|, |\mathbf{q}_2|)  & =  \bar{f}_{\nu l}(|\mathbf{q}_2|)  -  \bar{f}_{\nu H}(|\mathbf{q}_1|) \underset{\slashed{\rm qs}}{\to} \bar{f}_{\nu l} (|\mathbf{q}_2|);
	\end{aligned}
\end{equation}
and 
\begin{equation}
	\begin{aligned}
		q_{i\pm}^{(j)} =& \left| \frac{\epsilon_j \Delta(m_k,m_i,m_j) \pm |\mathbf{q}_j| \sqrt{\Delta^2(m_k,m_i,m_j)+4 m^2_i m^2_j}}{2 m^2_j} \right|
		\label{eq:intlimits_compact}
	\end{aligned}
\end{equation}
are the integration limits, with the understanding that $i \neq j \neq k$, and the mapping $(1,2,3) \leftrightarrow(\nu_H,\nu_l,\phi)$ applies.


\section{\texorpdfstring{$\ell \geq 2$}{ell geq 2} effective collision integrals}
\label{sec:collisionintegral}

We outline in this appendix the reduction of the $\ell \geq 2$ collision integrals~\eqref{eq:dpidt} to the form~\eqref{eq:effectiveEoM}.  Firstly, we note that the effective collision integral~\eqref{eq:dpidt} is a sum of three terms, which we label
\begin{equation}
\left(\frac{\mathrm{d} {\cal F}}{\mathrm{d}\tau}\right)_{C,\ell} \equiv 	\left(\frac{\mathrm{d} {\cal F}}{\mathrm{d}\tau}\right)_{C,\ell}^{(\nu H)} + 	\left(\frac{\mathrm{d} {\cal F}}{\mathrm{d}\tau}\right)_{C,\ell}^{(\nu l)}  + 	\left(\frac{\mathrm{d} {\cal F}}{\mathrm{d}\tau}\right)_{C,\ell}^{(\phi)} .
\end{equation}
Using the $\nu_H$ collision integral~\eqref{eq:C1_nuH} and setting $m_\phi=0$, the $\nu_H$ component evaluates straightforwardly to
\begin{equation}
\begin{aligned}
\left(\frac{\mathrm{d} {\cal F}}{\mathrm{d}\tau}\right)_{C,\ell}^{(\nu H)} = & \, a \, \tilde{\Gamma}_{\rm dec} \, \left( \frac{m_{\nu H}^2}{m_{\nu H}^2 - m_{\nu l}^2} \right)\,
T_0^{-4} \,  {\cal F}_\ell \int {\rm d} |\mathbf{q}_1|\; \frac{|\mathbf{q}_1|^{\ell+1}}{\epsilon_1^\ell} \, e^{-\epsilon_1 / T_0} \\
& \hspace{-0mm} \times \left\{  \int_{q_{2-}^{(\nu H)}}^{q_{2+}^{(\nu H)}} \mathrm{d}|\mathbf{q}_2| \,  \frac{|\mathbf{q}_2|}{\epsilon_2}  \, \left[ -\frac{|\mathbf{q}_1|^2}{\epsilon_1}+\frac{|\mathbf{q}_2|^2}{\epsilon_2} \, P_\ell(\cos \alpha^*) \right]+ \int^{q_{3+}^{(\nu H)}}_{q_{3-}^{(\nu H)}} \mathrm{d} |\mathbf{q}_3|  \, |\mathbf{q}_3| \, P_\ell(\cos \beta^*)  \right\} ,
\label{eq:rearrangenuH}
\end{aligned}
\end{equation}
where $\tilde{\Gamma}_{\rm dec}$ is defined in Eq.~\eqref{eq:tildegamma} and is approximately the boosted decay rate.

Observe that the $\ell$-dependence of the double integral~\eqref{eq:rearrangenuH} is contained entirely in (i)~$|\mathbf{q}_1|^{\ell}/\epsilon_1^\ell$, and (ii)~the Legendre polynomials.  Upon a small parameter expansion in $a^2 m_{\nu H}^2$ and  $a^2 m_{\nu l}^2$
to  ${\cal O}(a^4 m_{\nu H}^4,a^4 m_{\nu l}^4,a^4 m_{\nu H}^2 m_{\nu l}^2)$, the former becomes a quadratic polynomial in $\ell$:
\begin{equation}
\label{eq:q1expansion}
    \left(\frac{|\mathbf{q}_1|}{\epsilon_1}\right)^\ell \simeq 1 - \ell \frac{a^2 m_{\nu H}^2}{2 |\mathbf{q}_1|^2}+ \ell (\ell + 2) \frac{a^4 m_{\nu H}^4}{8 |\mathbf{q}_1|^4} + \cdots.
\end{equation}
For the benchmark $m_{\nu H} \simeq 0.2$~eV and a typical comoving momentum of $|\mathbf{q}_1| \simeq 3 T_{\nu,0} \simeq 5 \times 10^{-4}$~eV, this expansion is accurate at recombination ($z = z_{\rm rec} \simeq 1100$) to $0.2\%$ and $11$\% for $\ell=2$ and $\ell=10$, respectively, and improves to $0.03$\% and $1$\% at $z \simeq 1500$.  For the largest $m_{\nu H}$ value actually analysed in this work, i.e., $m_{\nu H} \simeq 0.1$~eV, sub-10\% accuracy is possible even for $\ell = 20$ at $z \simeq z_{\rm rec}$.
The small parameter expansion therefore appears well justified in the time frame of interest.

On the other hand, the arguments of the Legendre polynomials expand to
\begin{equation}
\begin{aligned}
\label{eq:arguments}
\cos \alpha^* &\simeq 1 + \frac{a^2 m_{\nu H}^2}{2 |\mathbf{q}_1|^2} \left(1 - \frac{|\mathbf{q}_1|}{|\mathbf{q}_2|} \right)+   \frac{a^2 m_{\nu l}^2}{2 |\mathbf{q}_2|^2} \left(1 - \frac{|\mathbf{q}_2|}{|\mathbf{q}_1|} \right)  + \frac{a^4 m_{\nu H}^2 m_{\nu l}^2}{4 |\mathbf{q}_1|^2 |\mathbf{q}_2|^2} - \frac{a^4 m_{\nu H}^4}{8 |\mathbf{q}_1|^4} - \frac{a^4 m_{\nu l}^4}{8 |\mathbf{q}_2|^4} +\cdots,\\
\cos\beta^* & \simeq 1 +\frac{a^2 m_{\nu H}^2}{2 |\mathbf{q}_1|^2} \left(1-\frac{|\mathbf{q}_1|}{|\mathbf{q}_3|} \right) + \frac{a^2 m_{\nu l}^2}{2 |\mathbf{q}_1| |\mathbf{q}_3|} - \frac{a^4 m_{\nu H}^4}{8 |\mathbf{q}_1|^4} + \cdots,
\end{aligned}
\end{equation}
while the Legendre polynomials themselves expand again to polynomials in $\ell$:
\begin{equation}
\label{eq:expandedlegendre}
P_\ell(1+x) \simeq 1 + \frac{1}{2} \ell (\ell+1) x + \frac{1}{16} (\ell -1) \ell (\ell + 1) (\ell + 2) x^2 + \cdots.
\end{equation}
Substituting Eq.~\eqref{eq:arguments} into Eq.~\eqref{eq:expandedlegendre} and keeping terms to 
${\cal O}(a^4 m_{\nu H}^4,a^4 m_{\nu l}^4,a^4 m_{\nu H}^2 m_{\nu l}^2)$, we immediately find in combination with Eq.~\eqref{eq:q1expansion} that the double integral~\eqref{eq:rearrangenuH} must be a quartic polynomial in $\ell$ with $\ell$-independent coefficients.  We shall not write out the polynomial in full here, but suffice it to say that  this is the origin of quartic $\ell$-dependence of the $\alpha_{\ell}$ coefficients~\eqref{eq:alpha}.  As with Eq.~\eqref{eq:q1expansion}, for the benchmark $m_{\nu H} \simeq 0.2$~eV and  typical comoving momenta $|\mathbf{q}_1| \simeq |\mathbf{q}_2|  \simeq 3 T_{\nu,0} \simeq 5 \times 10^{-4}$~eV, the small parameter expansion~\eqref{eq:expandedlegendre} is sub-percent accurate at $z \simeq z_{\rm rec}$ for $\ell \leq 5$ and incurs at most an error of $15$\%  for $\ell=10$.  It is thus a very good approximation.

Similarly to Eq.~\eqref{eq:rearrange2}, using the $\nu_l$ collision integral~\eqref{eq:C1_nul} under the assumption of $m_\phi=0$ yields
\begin{equation}
	\begin{aligned}
	\label{eq:fnul}
		\left(\frac{\mathrm{d} {\cal F}}{\mathrm{d}\tau}\right)_{C,\ell}^{(\nu l)} = & \, a \, \tilde{\Gamma}_{\rm dec} \,  \left( \frac{m_{\nu H}^2}{m_{\nu H}^2 - m_{\nu l}^2} \right)\, T_0^{-4} \frac{g_{\nu l}}{g_{\nu H}} {\cal F}_\ell \int {\rm d} |\mathbf{q}_2|\; \frac{|\mathbf{q}_2|^{\ell + 1}}{\epsilon_2^2} \\
	&\hspace{0mm} \times  \left\{ \int_{q_{1-}^{(\nu l)}}^{q_{1+}^{(\nu l)}}  \mathrm{d} |\mathbf{q}_1|   \frac{|\mathbf{q}_1|}{\epsilon_1}\left[ \frac{|\mathbf{q}_1|^2}{\epsilon_1}   P_\ell(\cos \alpha^*) - \frac{|\mathbf{q}_2|^2}{\epsilon_2} \right] e^{-\epsilon_1/T_0}  -  \int_{q_{3-}^{(\nu l)}}^{q_{3+}^{(\nu l)}}  \mathrm{d} |\mathbf{q}_3|  \,  |\mathbf{q}_3| \, P_\ell(\cos \gamma^*)  e^{-\epsilon_1/T_0} \right\} .
	\end{aligned}
\end{equation}
To progress further, following Ref.~\cite{Barenboim:2020vrr} we swap the order of integration, so that the final momentum integration to perform is always over $|\mathbf{q}_1| \in [0,\infty)$.  The swap between the $|\mathbf{q}_1|$ and  $|\mathbf{q}_2|$ double integral is straightforward, and entails setting the new integration limits as per
\begin{equation}
\int {\rm d} |\mathbf{q}_2| \int_{q_{1-}^{(\nu l)}}^{q_{1+}^{(\nu l)}} {\rm d} |\mathbf{q}_1| =  \int {\rm d} |\mathbf{q}_1| \int_{q_{2-}^{(\nu H)}}^{q_{2+}^{(\nu H)}} {\rm d} |\mathbf{q}_2|.
\end{equation}
To apply the same trick to the $|\mathbf{q}_2|$ and  $|\mathbf{q}_3|$ pair, however, we first need to change of the integration variable $|\mathbf{q}_3|$ to $|\mathbf{q}_1|$ via $|\mathbf{q}_3| \to \epsilon_1 - \epsilon_2$.  Then, recognising that integration limits can also be written as Heaviside functions, we can establish the relation~\cite{Barenboim:2020vrr} 
\begin{equation}
\begin{aligned}
\left. \Theta(|\mathbf{q}_3| - {q}_{3-}^{(\nu l)} ) \Theta({q}_{3+}^{(\nu l)}-|\mathbf{q}_3|)\right|_{|\mathbf{q}_3| \to \epsilon_1 - \epsilon_2} 
& = \left.  \Theta(1-\cos^2\gamma^*) \right|_{|\mathbf{q}_3| \to \epsilon_1 - \epsilon_2} \\
&  =  \Theta(1-\cos^2 \alpha^*) \\
& = \Theta(|\mathbf{q}_2| - {q}_{2-}^{(\nu H)} ) \Theta({q}_{2+}^{(\nu H)}-|\mathbf{q}_2|).
\label{eq:rearrange2}
\end{aligned}
\end{equation}
Note that, while we have assumed $m_\phi=0$ to simplify our calculations, the relation~\eqref{eq:rearrange2} follows from kinematics and holds for all mass values  {\it provided} we define the change of variable more generally as  $\epsilon_3 \to \epsilon_1 - \epsilon_2$.  Thus, Eq.~\eqref{eq:fnul} becomes
 \begin{equation}
 \begin{aligned}
 	\left(\frac{\mathrm{d} {\cal F}}{\mathrm{d}\tau}\right)_{C,\ell}^{(\nu l)} = & \, a \, \tilde{\Gamma}_{\rm dec} \, \left( \frac{m_{\nu H}^2}{m_{\nu H}^2 - m_{\nu l}^2} \right)\,  T_0^{-4} \,  {\cal F}_\ell \int {\rm d} |\mathbf{q}_1|\; \frac{|\mathbf{q}_1|^{\ell+1}}{\epsilon_1^\ell} \, e^{-\epsilon_1 / T_0} \\
 	&\hspace{0mm} \times  \int_{q_{2-}^{(\nu H)}}^{q_{2+}^{(\nu H)}} \mathrm{d}|\mathbf{q}_2|  \, \frac{\epsilon_1^{\ell-2}}{|{\bf q}_1|^{\ell-2}} \, \frac{|\mathbf{q}_2|^{\ell+1}}{\epsilon_2^\ell} \left[ P_\ell(\cos \alpha^*) 
 - \frac{|\mathbf{q}_2|^2 \epsilon_1}{\epsilon_2 |\mathbf{q}_1|^2}  - \frac{\epsilon_1}{|\mathbf{q}_1|^2} (\epsilon_1 - \epsilon_2) P_\ell(\cos \gamma^*)\big|_{|\mathbf{q}_3| \to \epsilon_1 - \epsilon_2} \right] ,
 \label{eq:rearrangenul}
 \end{aligned}
 \end{equation}
where we have used $g_{\nu l } = g_{\nu H} = 2$. Again, the term $|\mathbf{q}_2|^\ell/\epsilon_2^\ell$ can be expanded in the small parameters
$a^2 m_{\nu H}^2 $ and  $a^2 m_{\nu l}^2$ to ${\cal O}(a^4 m_{\nu H}^4,a^4 m_{\nu l}^4,a^4 m_{\nu H}^2 m_{\nu l}^2)$ as per Eq.~\eqref{eq:q1expansion}  save for the replacements $m_1 \to m_2$ and $|\mathbf{q}_1| \to |\mathbf{q}_2|$, while the argument of the second Legendre polynomial expands to
\begin{equation}
\cos\gamma^* \simeq 1 - \frac{a^2 m_{\nu H}^2}{2 |\mathbf{q}_2| |\mathbf{q}_3|}+
\frac{a^2 m_{\nu l}^2}{2 |\mathbf{q}_2|^2} \left(1+\frac{|\mathbf{q}_2|}{|\mathbf{q}_3|} \right)  - \frac{a^4 m_{\nu l}^4}{8 |\mathbf{q}_2|^4} + \cdots.
\end{equation}
Then, by Eq.~\eqref{eq:expandedlegendre} we can again expect the $\ell$-dependence of the double integral~\eqref{eq:rearrangenul} to simplify to that of 
a quartic polynomial in $\ell$.

Finally, the $\phi$ component of the effective collision integral can be similarly established from the $\phi$ collision integral~\eqref{eq:C1_phi} following the same procedure as above:
\begin{equation}
\begin{aligned}
		\left(\frac{\mathrm{d} {\cal F}}{\mathrm{d}\tau}\right)_{C,\ell}^{(\phi)} = & \,  a \, \tilde{\Gamma}_{\rm dec} \,  \left( \frac{m_{\nu H}^2}{m_{\nu H}^2 - m_{\nu l}^2} \right)\, 
		T_0^{-4} \,  {\cal F}_\ell \int {\rm d} |\mathbf{q}_1|\; \frac{|\mathbf{q}_1|^{\ell+1}}{\epsilon_1^\ell} \, e^{-\epsilon_1 / T_0} \\
		& \hspace{-7mm} \times \int_{q_{3-}^{(\nu H)}}^{q_{3+}^{(\nu H)}} \mathrm{d}|\mathbf{q}_3|   \frac{\epsilon_1^{\ell-2}}{|{\bf q}_1|^{\ell-2}}   |\mathbf{q}_3| \bigg[P_\ell(\cos \beta^*)  - \frac{|\mathbf{q}_3| \epsilon_1}{|\mathbf{q}_1|^2}  
	- \frac{\epsilon_1}{|\mathbf{q}_1|^2} \frac{(\epsilon_1 - |\mathbf{q}_3|)^2 - a^2 m_{\nu l}^2}{(\epsilon_1 - |\mathbf{q}_3|)} P_\ell(\cos \gamma^*)\big|_{\epsilon_2 \to \epsilon_1 - |\mathbf{q}_3|} \bigg] ,
		\label{eq:rerrangephi}
\end{aligned}
\end{equation}
with $g_\phi=1$ as an input.   A quartic polynomial in $\ell$ again follows from a small parameter expansion in $a^2 m_{\nu H}^2$ and  $a^2 m_{\nu l}^2$ to ${\cal O}(a^4 m_{\nu H}^4,a^4 m_{\nu l}^4, a^4 m_{\nu H}^2 m_{\nu l}^2)$.

Then, collecting all terms~\eqref{eq:rearrangenuH}, \eqref{eq:rearrangenul}, and \eqref{eq:rerrangephi} yields the effective collision integral~\eqref{eq:effectiveEoM}.


\section{A revised random walk picture of isotropisation}
\label{sec:heuristic}

\subsection{Preliminaries}

Consider the process $\nu_H \to \nu_l + \phi$ in the rest-frame of $\nu_H$, which we denote ${\cal S}'$.
  Suppose the decay products are emitted along the $y$-axis.  Assuming $m_\phi=0$, the decay products have momentum
\begin{equation}
\label{eq:pstar}
p^* = \frac{ m_{\nu H}^2-m_{\nu l}^2}{2 m_{\nu H}} \equiv \frac{\Delta m_\nu^2}{2 m_{\nu H}}.
\end{equation}  
Boosting the system in the $x$-direction to the laboratory frame ${\cal S}$ at an ultra-relativistic speed $v_{\nu H}$, we find that the decay product~$i$ has momentum components in the $x$- and $y$-directions given respectively by
\begin{equation}
\begin{aligned}
\label{eq:momentum}
p^{(i)}_{x} & = \gamma_{\nu H} v_{\nu H} \sqrt{{p^*}^2+m_i^2}, \\
p^{(i)}_{y} & =\pm p^*,
\end{aligned}
\end{equation}
where $\gamma_{\nu H} \equiv (1-v_{\nu H}^2)^{-1/2} = E_{\nu H}/m_{\nu H}$ is the Lorentz factor, and $\tan \theta_i = p_y^{(i)}/p^{(i)}_x$ gives the angle of the decay trajectory relative to the initial direction of $\nu_H$.  For the massless $\phi$ particle, this angle always evaluates $|\theta_ \phi| \simeq \gamma_{\nu H}^{-1}$ in the limit $\gamma_{\nu H} \gg 1$, while for a massive daughter neutrino $\nu_l$ we find a suppressed emission angle of $|\theta_{\nu l}| \simeq [\Delta m_\nu^2/(2 m_{\nu H}^2)] \, \gamma_{\nu H}^{-1}$ in the same limit.  The latter, or more directly, $p^*$ in Eq.~\eqref{eq:pstar}, is the origin of the new phase space factor~\eqref{eq:phasespace} in the effective transport rate~\eqref{eq:effectiveEoM}.

In the case of {\it two massless} decay products, Ref.~\cite{Hannestad:2005ex} argues that, because of the small decay opening angle, $\theta = \theta_\phi = \theta_{\nu l} \simeq \gamma_{\nu H}^{-1} \ll 1$, it takes $N \simeq 1/\theta^2$  number of decay/inverse decay events to turn a momentum vector by 90 degrees.  This estimate comes from a standard random walk argument generalised to momentum space, where the random walker has a mean square displacement of $\sim (p_{\nu H} \theta_i)^2$ and {\it equal probabilities} of going $+|\theta|$ or $-|\theta|$ after each decay or inverse decay event, while maintaining its absolute momentum $p_{\nu H}$.
Assuming further that each event occurs over a timescale of $\sim 1/\Gamma_{\rm dec}$, where $\Gamma_{\rm dec} = \gamma_{\nu H}^{-1} \Gamma_{\rm dec}^0$ is the boosted decay rate, one immediately establishes that the time required to execute this 90-degree turn must be 
\begin{equation}
\label{eq:t2}
T_2 \sim \theta^{-2} \gamma/\Gamma_{\rm dec}^0.
\end{equation}
Reference~\cite{Hannestad:2005ex} identifies $T_2$ with the timescale over which the quadrupole of the $(\nu_H,\nu_l,\phi)$-system is lost through decay/inverse decay, in drastic contrast with the much longer isotropisation timescale,
\begin{equation}
\label{eq:t4}
T_4 \sim  \theta^{-4}\gamma/\Gamma_{\rm dec}^0,
\end{equation}
established from our relaxation time/linear response calculation in the  same $m_{\nu l}=0$ limit~\cite{Barenboim:2020vrr}. 

It is not completely clear how one might generalise $T_2$ in Eq.~\eqref{eq:t2} to the case of a massive $\nu_l$, since the two decay products must now be emitted at different angles.  One possibility might be to identify the emission angle~$\theta$  with the geometric mean,
$\theta \equiv \pm \sqrt{|\theta_{\nu l} \theta_\phi|}$.  At minimum, this choice of $\theta$  yields a $T_4$ from Eq.~\eqref{eq:t4} to within ${\cal O}(1)$ factors of the timescale obtained via rigorous calculations in this work [Eq.~\eqref{eq:effectiveEoM}].  This identification also appears to be the only sensible choice if we must boil what is in principle a two-parameter problem (for two decay products) down to a description in terms of one angle $\theta$ alone.
Our task in this appendix, therefore, is to reconcile the two timescales~$T_2$ and~$T_4$, and to supply a random walk interpretation for the latter in terms of $\theta$.


\subsection{Coverage versus isotropisation}

To reconcile the two timescales $T_2$ and $T_4$, and to understand why it is the latter, not the former, that gives the isotropisation timescale, we observe first of all that as a representation of isotropisation of an anisotropic system via scattering, the random walk picture as presented above in fact contains a number of inconsistencies.  For one, the same reasoning would lead us to conclude that the momentum vector of a single particle can be turned by 180 degrees in time $\sim 4 \times T_2$. This is in itself not an improbable outcome.  It makes no physical sense, however, to associate this outcome with the timescale over which the {\it dipole} of the $(\nu_H,\nu_l,\phi)$-system is lost through decay/inverse decays, because as an isolated system, momentum conservation always preserves the system's dipole.  This simple  consideration illustrates that one needs to be extremely cautious when identifying the behaviour of {\it one single particle} with changes to the {\it bulk properties} of a system.  They are {\it a priori} not the same thing.

Next, we note that $\nu_H$ decay always happens and isotropisation is limited only by inverse decay.  However, in writing down $\Gamma_{\rm dec}$ as a universal {\it inverse} decay rate without qualification, there is an implicit assumption that the scattering centres effecting the inverse decay are themselves isotropic in momentum space, such that the probability of $\nu_H$ being emitted in the $+|\theta|$ or $-|\theta|$ direction is everywhere the same.  This assumption is not true for our $(\nu_H,\nu_l,\phi)$-system, which is constantly subject to tidal gravitational forces that generate ${\cal O}(10^{-5})$ local momentum anisotropies.  These anisotropies give rise to anisotropic probabilities in the emission direction of $\nu_H$, which in turn bias the random walk. At face value this bias is tiny---of order $10^{-5}$.   However, we must keep in mind that it has the same physical origin as the momentum anisotropy we wish to wipe by inverse decay, i.e., the bias and the momentum anisotropy are of {\it exactly the same order of magnitude}.  A consistent picture of isotropisation therefore {\it cannot} ignore the bias, and understanding how it influences the random walks is key to explaining why the isotropisation timescale is $T_4$, not~$T_2$.

\begin{figure*}[t]
    \centering
    \includegraphics[width=\textwidth]{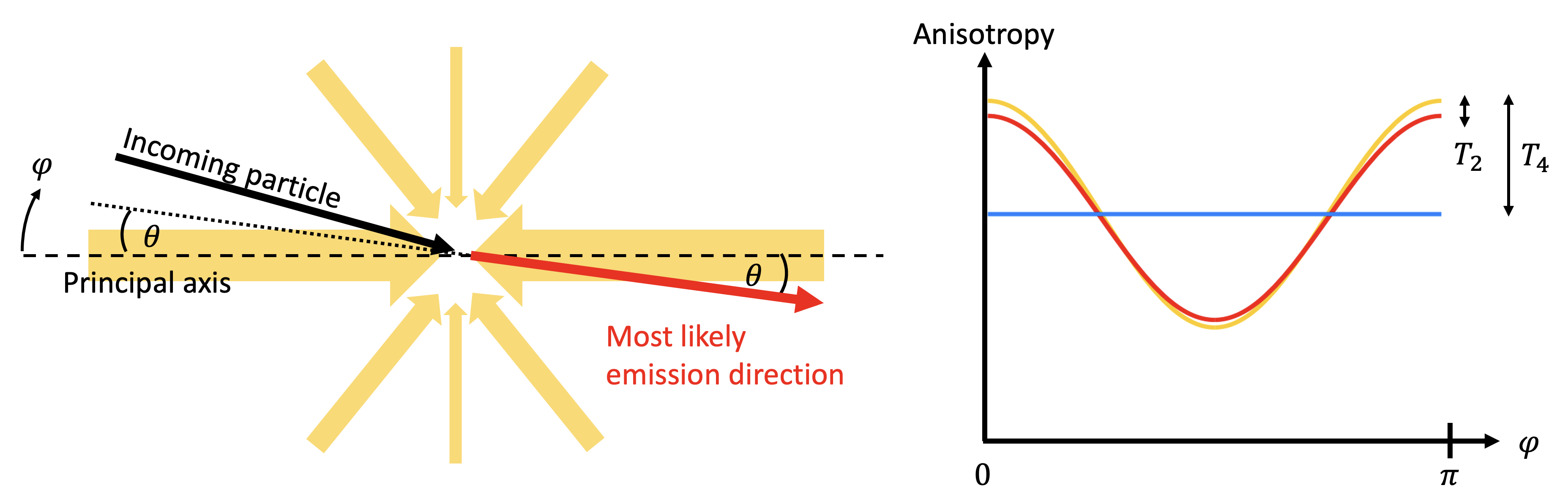}
    \caption{Schematic illustrating how a quadrupole anisotropy in the background population can enhance the probability of $\nu_H$ being emitted in a direction as aligned as possible with the principal axis of anisotropy. Observe in this example that the emitted $\nu_H$ is misaligned with the principal axis of anisotropy by an angle $\sim \theta$.  This small misalignment leads to a distribution of random walkers after one coverage time~$T_2$ [Eq.~\eqref{eq:t2}] that is slightly dispersed  relative to the distribution of the background population, i.e., the walker population (red line in the right plot) is slightly isotropised relative to the background population (yellow line) after one coverage time~$T_2$.  Then, by repeatedly updating the background anisotropy with the new walker population anisotropy (i.e., we
     ``swap'' the roles of the walker and the background populations) after every $T_2$, we will eventually after $M \sim 1/\theta^{2}$ updates end up with a system that is completely isotropised (blue line).  The total time this isotropisation process takes is $T_4 \sim M \times T_2$ [Eq.~\eqref{eq:t4}]. \label{fig:anisotropy}} 
\end{figure*}

To simplify the discussion, let us separate the system into a population of ``walkers'' and a population of ``background'' particles that recombine with the walkers to form $\nu_H$.
As illustrated in Fig.~\ref{fig:anisotropy}, when a walker encounters a background with a quadrupole anisotropy,
 strict angular and momentum requirements for an inverse decay to happen favour the resulting $\nu_H$ to be emitted in a direction as closely aligned as possible with the principal axis of anisotropy.  The alignment is however never exact, as $\nu_H$ is always emitted at an angle $\sim \theta$ relative to both the trajectory of the walker and of the background particle that combines with it.
In other words, we can think of the random walkers as ``samplers'' of an ``emission direction function'' $\mathscr{E}$ that has a gradient in the angular direction  following the anisotropy of the background, but with a small isotropic dispersion  over the emission angle~$\theta$---isotropic because the inverse decay interaction itself is invariant under rotation; any asymmetry in the $\nu_H$ emission direction is due solely to the background anisotropies.
  This small isotropic dispersion is important, as it is precisely this dispersion that allows the $\ell\geq 2$ anisotropies of the system to be lost through decay/inverse decay, while preserving the dipole.

Now, as mentioned above, anisotropies in the cosmological context are typically no more than $\sim 10^{-5}$.  We can therefore still expect a random walker to cover all angular directions in  $N \sim 1/\theta^2$ steps, or, equivalently, in a time~$T_2$: for this reason we shall call $T_2$ the {\it coverage time}. It is also over this timescale that the random walker ``forgets'' where it has come from.
However, if we were to ask, what would happen to a whole population of random walkers after one coverage time~$T_2$, then the answer would be we would find a walker population with no memory of its initial configuration but that is now distributed in a way that mimics the emission direction function $\mathscr{E}$,  itself dictated by the underlying anisotropy of the background.  Thus, we have not managed to wipe out the system's anisotropy over time~$T_2$; we have merely sampled the emission direction function $\mathscr{E}$ and hence turned the random walker population into a slightly less anisotropic version of the background.
At this point the reader may find our description reminiscent of a Markov Chain. It is indeed a Markov Chain, where the emission direction function $\mathscr{E}$ plays the role of the likelihood function, and $T_2$ is the burn-in time.

We can easily extend this picture to include updates to the emission direction function $\mathscr{E}$ as the system loses its anisotropy, in order to estimate the isotropisation timescale, i.e., the time it takes to flatten $\mathscr{E}$.
In reality the update must of course take place continuously, as the loss of anisotropy is itself a continuous process.
In our discretised picture with one walker and one background population, however, we can mimic the update by swapping the roles of the walkers and the background after every coverage time $T_2$. That is, at every swap the new emission direction function $\mathscr{E}_{\rm new}$ is an isotropically smeared version of the old function $\mathscr{E}_{\rm old}$ over the emission angle~$\theta$ given by
\begin{equation}
\mathscr{E}_{\rm new}(\varphi) = \int {\rm d} \varphi' \, \mathscr{E}_{\rm old} (\varphi') f(\varphi'-\varphi, \theta),
\end{equation}
where $f(\varphi,\theta)$ is a normalised filter function of width $\sim \theta$. Then, for the quadrupole anisotropy of Fig.~\ref{fig:anisotropy}---given by $\mathscr{E}_{\rm old} = (3 \cos^2 \varphi-1)/2$---and assuming a  top-hat filter $f(\varphi,\theta)= \Theta(\varphi+\theta)\Theta(\theta-\varphi)/(2 \theta)$, we see immediately that at its maximum point $\mathscr{E}_{\rm old}(\varphi=0)=1$, the new emission direction function $\mathscr{E}_{\rm new} (\varphi=0)$ must reduce to
\begin{equation}
\begin{aligned}
\mathscr{E}_{\rm new}(\varphi=0) & = \frac{1}{4} + \frac{3 \sin(2 \theta)}{8 \theta} \\
& \simeq 1 - \frac{\theta^2}{2}, \quad \theta \ll 1
\end{aligned}
\end{equation}
after the first swap, i.e., it reduces by an amount $\theta^2/2$; had we used a normalised Gaussian filter of width $\theta$, we would have obtained a similar reduction of $3\theta^2/2$.  Thus, we arrive at the conclusion that the number of swaps required to flatten $\mathscr{E}$ must be $M \sim 1/\theta^2$.  Then, together with the coverage timescale $T_2$ from Eq.~\eqref{eq:t2}, we find an isotropisation timescale of $\sim M \times T_2 \sim \theta^{-4} \gamma/\Gamma_{\rm dec}^0$, which  is precisely $T_4$ of Eq.~\eqref{eq:t4}.

In conclusion, the random walk argument employed in Ref.~\cite{Hannestad:2005ex} to support an isotropisation timescale of $T_2$ [Eq.~\eqref{eq:t2}] carries a strong but unjustified assumption that the background on which a random walker scatters is {\it already isotropised}.  Once this assumption is dropped, $T_2$ is no longer sufficient time for isotropisation, and should be reinterpreted as the coverage time for a random walker to survey the system's configuration once and to forget its previous state,
while the anisotropy decreases only by a small amount $\theta^2$.
The true isotropisation timescale is a much longer $T_4$ [Eq.~\eqref{eq:t4}], stemming from the need for each random walker to survey and forget its own previous state another $M \sim 1/\theta^2$ times, before all walkers will agree on their final configurations.


\section{Parameter estimation}
\label{sec:parameter}

\begin{table*}[t]
	\begin{center}
	\begin{tabular}{c|ccc}
		\toprule
		&		\multicolumn{3}{c}{\rotatebox[origin=c]{0}{\bf Scenario A}} \\
		\cline{2-4}
    & $X=52$ & $X=300$ & $X=595$  \\
		\toprule
	$\omega_b$ &$0.02249 \pm 0.00016$ & $0.02248 \pm 0.00016$&$0.02249\pm 0.00016$\\
	$\omega_c $ &$0.1208 \pm 0.0013$&$0.1208 \pm 0.0013$ & $0.1208 \pm 0.0012$ \\
	$100 \theta_s$ &$1.04178 \pm 0.00029$&$1.04181 \pm 0.00030$ & $1.04179 \pm 0.00030$\\
	$\tau_{\rm reio}$ &$0.0561 \pm 0.0074$ &$0.0560 \pm 0.0080$ & $0.0567 \pm 0.00755$\\
	$n_s$ &$0.9747 \pm 0.0045$ &$0.9751 \pm 0.0049$ & $0.9759 \pm 0.0049$\\
	$\ln (10^{10} A_s)$ &$3.0424 \pm 0.0145$ &$3.0431 \pm 0.0154$ & $3.0443 \pm 0.0149$\\
	$Y (s^{-1})$ & $< 18611$ &$< 41636$ & $< 74448$ \\
	\cline{1-4}
	$h$ &$0.6768 \pm 0.0557$ &$0.6768 \pm 0.0568$ & $0.6768 \pm 0.0538$\\
	$\sigma_8$ &$0.8276 \pm 0.0063$ &$0.8280 \pm 0.0064$ & $0.8287 \pm 0.0062$\\
		\toprule
		&		\multicolumn{2}{c}{\rotatebox[origin=c]{0}{\bf Scenario B}} & 
		\parbox[t]{15mm}{\multirow{2}{*}{\rotatebox[origin=c]{0}{{\bf $\Lambda$CDM}~\cite{Planck:2018vyg}}}}\\
		\cline{2-3}
    & $X=298$ & $X=595$ &  \\
		\toprule
	$\omega_b$ &$0.02229 \pm 0.00015$&$0.02227 \pm 0.00015$ &$0.02237 \pm 0.00015$\\
	$\omega_c $ & $0.1200 \pm 0.00122$&$0.1201 \pm 0.0012$& $0.1200 \pm 0.0012$ \\
	$100 \theta_s$ & $1.04188 \pm 0.00030$ &$1.04191 \pm 0.00029 $ &  $1.04092 \pm 0.00031$\\
	$\tau_{\rm reio}$ & $0.0545 \pm 0.0072$ &$0.0539 \pm 0.0072$ & $0.0544\pm 0.0073$\\
	$n_s$ & $0.9690 \pm 0.0044$&$0.9674 \pm 0.0043$& $0.9649 \pm 0.0042$ \\
	$\ln (10^{10} A_s)$ & $3.0424 \pm 0.0138$ &$3.0423 \pm 0.0140 $& $3.044 \pm 0.014$\\
	$Y (s^{-1})$ & $<2286$ & $< 2288$  & -- \\
	\cline{1-4}
	$h$ & $0.6780 \pm 0.5592$ &$0.6778 \pm 0.5592$ & $0.6736 \pm 0.0054$ \\
	$\sigma_8$ & $0.8245 \pm 0.0060$ &$0.8243 \pm 0.0060$& $0.8111 \pm 0.0060$ \\
		\toprule
	\end{tabular}
\end{center}
\caption{Mean and 1D marginalised 68\% credible intervals for the base $\Lambda$CDM parameters and the marginalised one-sided 95\% credible limits on the effective transport rate~$Y$ derived from the Planck 2018 CMB data combination
TTTEEE+lowE+lensing~\cite{Planck:2018vyg}, for fixed values of the effective mass parameter $X$ in scenarios A and B.  The vanilla $\Lambda$CDM estimates are taken from Ref.~\cite{Planck:2018vyg}.\label{tab:allconstraints}}
\end{table*}

\begin{figure}[p]
    \centering
    \includegraphics[width=\textwidth]{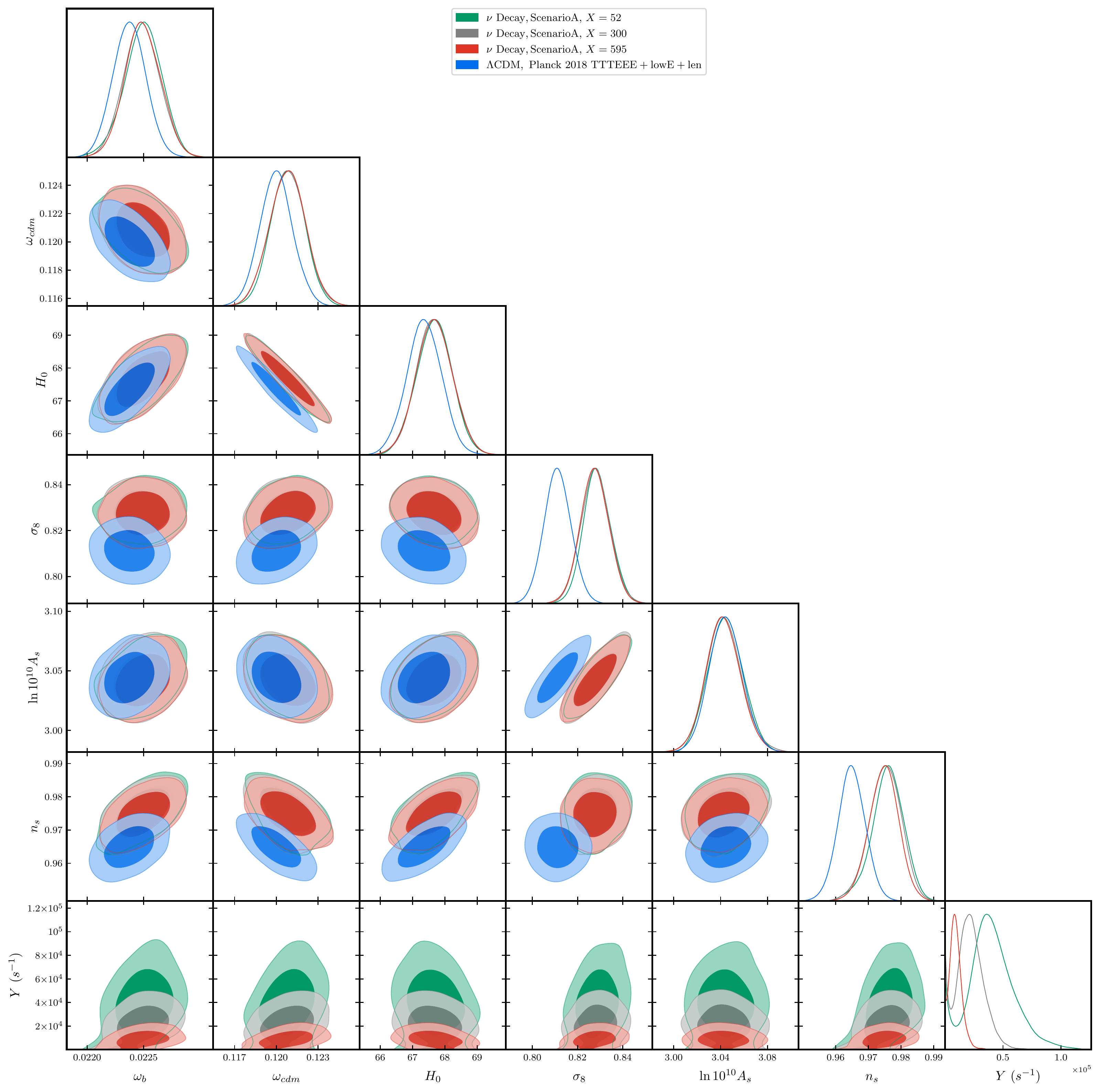}
    \caption{2D marginalised 68\% and 95\% contours for the variables of Scenario A, for fixed values of the effective mass parameter $X$.  We show also the corresponding contours from a vanilla $\Lambda$CDM fit.}
    \label{fig:my_labelA}
\end{figure}

\begin{figure}[p]
    \centering
    \includegraphics[width=\textwidth]{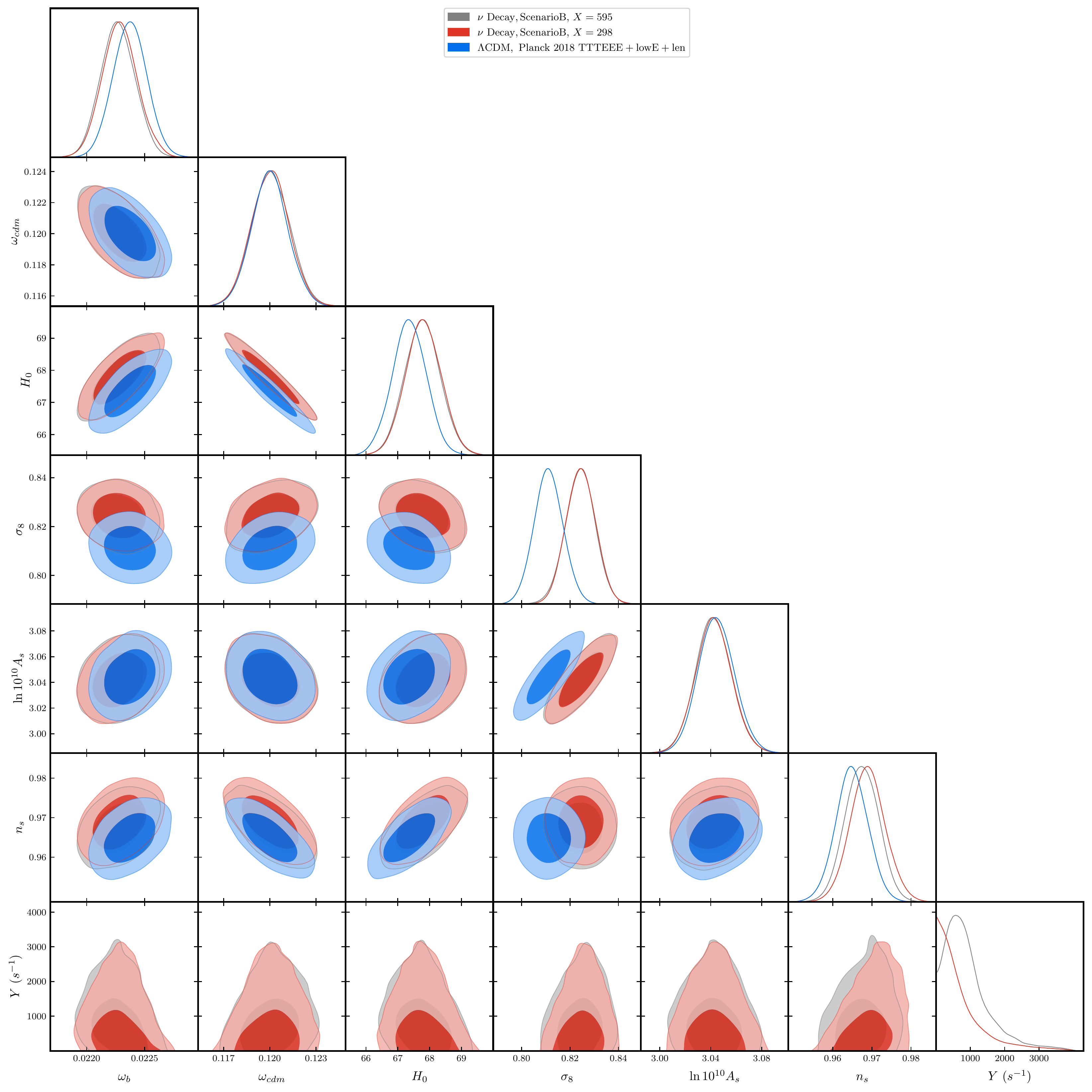}
    \caption{Same as Fig.~\ref{fig:my_labelA}, but for scenario B.}
    \label{fig:my_labelB}
\end{figure}

We provide in this section summary statistics of our Markov Chains from Sect.~\ref{sec:data}.  Table~\ref{tab:allconstraints} shows the mean and 1D marginalised credible intervals for the variables of Eqs.~\eqref{eq:fitparamsA}  (scenario A) and~\eqref{eq:fitparamsB} (scenario B),  as well as for two derived parameters, the reduced Hubble expansion rate $h$ and the RMS fluctuation on $8$~Mpc $\sigma_8$, for all choices of fixed $X=X_{\rm scan}$ considered. For comparison we also provide the parameter estimates in a vanilla $\Lambda$CDM fit to the same data combination  taken from Ref.~\cite{Planck:2018vyg}.  Figures~\ref{fig:my_labelA} and~\ref{fig:my_labelB} show the  corresponding 2D marginalised contours.

\end{document}